\documentclass[lettersize,journal,x11names,compsoc]{IEEEtran}
\usepackage{amsmath,amsfonts,amsthm}
\usepackage{enumitem}
\usepackage{todonotes}
\usepackage{dsfont}
\usepackage{cuted}
\usepackage{booktabs}
\usepackage{algorithm}
\usepackage{algpseudocode}
\usepackage{array}
\usepackage{subcaption}
\usepackage{textcomp}
\usepackage{xcolor}
\usepackage{stfloats}
\usepackage{url}
\usepackage{verbatim}
\usepackage{graphicx}
\usepackage{tabularx}
 \usepackage{multirow}
\usepackage{cite}
\usepackage{tikz}
\usepackage{gnuplot-lua-tikz}
\usepackage{pgfplots}
\usetikzlibrary{calc,math}
\usetikzlibrary{angles, intersections}
\pgfplotsset{compat=1.9}
\usepgfplotslibrary{fillbetween}
\usetikzlibrary{external}
\newtheorem{problem}{Problem}

\newcommand{\pelao}[3]{
    \draw[
        evaluate={
            \x = (#1); 
            \y = (#2); 
            \r = (#3); 
            \xl = \x-\r; 
            \xr = \x+\r; 
            \ya = \y-\r; 
            \yb = \ya-.75*\r; 
            \yc = \yb-\r; 
            \ybd = \yb-0*\r; 
            \ycd = \yc-0.75*\r; 
        }
    ]
    (\x, \y) circle (\r) 
    (\x, \ya) -- (\x,\yc) 
    (\xl, \ybd) -- (\x, \yb) -- (\xr, \ybd) 
    (\xl, \ycd) -- (\x, \yc) -- (\xr, \ycd) 
    ; %
} 

\newcommand{\runpelao}[4]{
    \draw[
        rotate around={(#4):((#1,#2))},
        evaluate={
            \xn = (#1); 
            \yn = (#2); 
            \r = (#3); 
            \xnl = \xn-\r; 
            \xnr = \xn+\r; 
            \yna = \yn-\r; 
            \ynb = \yna-.75*\r; 
            \ync = \ynb-\r; 
            \ynbd = \ynb-0*\r; 
            \yncd = \ync-0.75*\r; 
        }
    ]
    (\xn, \yn) circle (\r) 
    (\xn, \yna) -- (\xn,\ync) 
    (\xnl, \ynbd) -- (\xn, \ynb) -- (\xnr, \ynbd) 
    (\xnl, \yncd) -- (\xn, \ync) -- (\xnr, \yncd) 
    (\xnr, \ynbd) -- (\xnr, \ynbd+.1) 
    (\xnl, \ynbd) -- (\xnl, \ynbd-.1) 
    ; %
} 

\usepackage{amsmath}

\DeclareMathOperator*{\argmin}{arg\,min}

\newtheorem{theorem}{Theorem}
\newtheorem{lemma}[theorem]{Lemma}

\begin{document}

\title{Towards Cooperative VRUs: Optimal Positioning Sampling for Pedestrian Awareness Messages }

\author{Jorge Martín-Pérez, Oscar Amador,
Markus Rydeberg, Linnéa Olsson, 
Alexey Vinel
\IEEEcompsocitemizethanks{%
\IEEEcompsocthanksitem{
Jorge Martín-Pérez is with Departamento de Ingeniería de Sistemas Telemáticos, ETSI Telecomunicación, Universidad Politécnica de Madrid, 28040, Spain. Email: jorge.martin.perez@upm.es.}
\IEEEcompsocthanksitem{Oscar Amador, Markus Rydeberg, and Linnéa Olsson are with the School of Information Technology, Halmstad University, 30118 Halmstad, Sweden (e-mail:oscar.molina@hh.se, \{marryd18,linols18\}@student.hh.se}
\IEEEcompsocthanksitem{Alexey Vinel is with the Karlsruhe Institute of Technology, 76131 Karlsruhe, Germany}
}
\thanks{This work was partially supported by Swedish Transport Administration in the project ``Here I go - avancerade funktioner för VRU medvetenhetsprotokoll'' TRV2023/104170, the Knowledge Foundation in the project “SafeSmart – Safety of Connected Intelligent Vehicles in Smart Cities” (2019-2024), the ELLIIT Strategic Research Network in the project ``6G wireless'' – sub-project ``vehicular communications'', the Helmholtz Program ``Engineering Digital Futures'',
the Remote Driver project (TSI-065100-
2022-003) funded by Spanish Ministry of Economic Affairs and Digital
Transformation, and the ECTICS project (PID2019-105257RB-C21), funded
by the Spanish Ministry of Economy and Competitiveness, and the Spanish
Ministry of Science and Innovation
}
\thanks{We thank professor Bernhard Sick for his suggestions
and fruitful discussions.}
\thanks{Manuscript received Month DD, 2023; revised Month DD, 2023.}
\thanks{©2023 IEEE. Personal use of this material is permitted. Permission
from IEEE must be obtained for all other uses, in any current or future
media, including reprinting/republishing this material for advertising
or promotional purposes, creating new or redistribution to servers or
lists, or reuse of any copyrighted component of this work in other
work}}

\markboth{IEEE Transactions on XXXX,~Vol.~XX, No.~X, Month~YYYY}%
{Shell \MakeLowercase{\textit{et al.}}: A Sample Article Using IEEEtran.cls for IEEE Journals}


\IEEEtitleabstractindextext{%
\begin{abstract}
Road safety is the main motivation for Cooperative Intelligent Transport Systems (C-ITS) in general, and vehicular communications (V2X) technology in particular. The V2X-based Vulnerable Road User (VRU) protection is an approach that relies on the persistent broadcasting of "beacon" awareness messages by a VRU mobile device. To this end the European Telecommunications Standards Institute (ETSI) has specified the Vulnerable Road User Awareness Message (VAM) as well as the overall ITS-G5 protocol stack enabling a variety of the V2X applications. 
This article studies how 
often pedestrians (a type of VRU) should
check their position
to issue a VAM. To that end, we characterize
the rate at which pedestrians generate
VAMs leveraging a recognized mobility model,
and formulate an optimization problem
to minimize the time elapsed between
VAMs. We propose an algorithm to solve the
problem in 802.11p and assess its accuracy
through numerical and simulation campaigns.
Results evidence the accuracy of our
VAM rate characterization, and evidence
that we decrease
ETSI positioning sampling rate by more
than 30\%. On top, 
our solution decreases the time between
VAMs, and increases the packet delivery
ratio.
In other words, our approach 
increases the pedestrians safety while
reducing the battery consumption of mobile devices.

\end{abstract}
\begin{IEEEkeywords}
Cooperative Intelligent Transport Systems (C-ITS), Cooperative Vehicles, Automated Driving, Vehicular Communications (V2X), Pedestrian-to-Anything (P2X), Road Safety, Vulnerable Road Users (VRU).
\end{IEEEkeywords}
}

\maketitle

\section{Introduction}

\IEEEPARstart{O}{ver} 270,000 pedestrians die every year in traffic accidents worldwide~\cite{Who2022}. In Europe alone, 19,897 people died in road accidents in 2021~\cite{Europe2023}. Pedestrians and cyclists ---part of the Vulnerable Road User (VRU) category--- make up 27\% of those fatalities. These numbers have prompted initiatives such as Vision Zero~\cite{ecVisionZero}, adopted by the European Commission (EC), which aims at reducing road fatalities and serious injuries to zero by 2050. Furthermore,
the United Nations
aim to halve the number
of deaths/injuries
from road traffic
accidents by 2030,
as specified
in~\cite[Target~3.6]{Agenda2030} of
the Sustainable
Development Goals.

Technology is one of the fronts that pave the road to Vision Zero. The development of Connected, Cooperative, and Automated Mobility (CCAM) is one of the cornerstones of the technological effort, and VRU protection is one of the main objectives of the industry and research communities working in CCAM. Vehicular communications (V2X) being a part of Cooperative Intelligent Transport Systems (C-ITS) for a plethora of the VRUs types (pedestrians, cyclists, etc.) are central in CCAM framework.  

We can divide VRU protection into two main categories: \textit{passive} (when vehicles or the infrastructure detect VRUs) and \textit{active} (when the VRU informs other road users of its presence). In the context of one specific type of the VRU, namely pedestrians, the latter approach is known as Pedestrian-to-Anything (P2X) communication. For an broader overview of the P2X systems, the reader is referred to the most recent survey~\cite{Bruno2023}. 

For P2X, standardization bodies such as Society of Automotive Engineers (SAE) International and the European Telecommunications Standards Institute (ETSI) have specified two services powered by beacon messages: the SAE Personal Safety Message (PSM)~\cite{SAEpsm}, and the ETSI Vulnerable Road User Awareness Message (VAM)~\cite{etsi-va}. There is already some work in the literature assessing the efficiency of PSM~\cite{IslamPSM} and VAM~\cite{combine,VAMv2xField,v2p-cv2x,assessvru}. The congestion in a broadcast communication channel, i.e., when many nodes (e.g., VRUs, vehicles) contend for the medium has a non-negligible effect on the beacon messages delivery. It is already exhibited in works related to vehicles broadcasting Cooperative Awareness Messages (CAMs)~\cite{Lyamin2017}\cite{OscarGOT}. Congestion effects can be expected to become even more prominent in case of pedestrians sending
VAMs due to their potentially higher density in comparison to vehicles. 


ETSI VAMs may congest the channel upon
high VRU density or large VAM generation
rates.
According to ETSI~\cite{etsi-va}, pedestrians
and other VRUs should generate a VAM if
there is a change of
position, heading, or speed. In particular,
VRUs must sample their position
at a rate above 10\,\textrm{Hz} to check if a VAM
must be generated~\cite{etsi-va}.

Despite the existing literature assessing
the performance of VRU VAMs
\cite{combine,VAMv2xField,v2p-cv2x,assessvru},
to the best of our knowledge there exists
only one work evaluating ETSI VA
parameters regarding the generation of
VAMs. In particular,
\cite{cluster-eval} evaluates how ETSI
parameters impact the performance of
VAMs within VRU clusters. However,
still there is no study on how the positioning
sampling impacts the
rate at which
VAMs are successfully
received.

\begin{figure*}[t]
    \begin{subfigure}[b]{0.28\textwidth}
        \centering
        \begin{tikzpicture}
    \def\D{1.5}
    \def\xzero{.4}
    \def\yzero{.7}

    \draw[|->] (0,0) -- (\xzero+2.5*\D,0)
        node[anchor=west] {$x$};

    \def\xpone{\xzero}
    \def\xptwo{\xzero+\D}
    \def\xpthree{\xzero+2*\D}
    \def\xpfour{\xzero+2*\D}
    \def\cabeza{.15}

    \pelao{\xpone}{\yzero}{\cabeza}
    \pelao{\xptwo}{\yzero}{\cabeza}
    \pelao{\xpthree}{\yzero}{\cabeza}

    \draw[|-|] (\xpone+.3,.5*\yzero)
        -- (\xptwo-.3,.5*\yzero)
        node[midway,above] {$\Delta$};
    \draw[|-|] (\xptwo+.3,.5*\yzero)
        -- (\xpthree-.3,.5*\yzero)
        node[midway,above] {$\Delta$};

    \def\arrlen{.35}
    \def\arrend{-.1}

    \draw[->] (\xpone-.1, \arrend-\arrlen)
        -- (\xpone-.1, \arrend);
    \draw[->] (\xptwo-.1, \arrend-\arrlen)
        -- (\xptwo-.1, \arrend);
    \draw[->] (\xpthree-.1, \arrend-\arrlen)
        -- (\xpthree-.1, \arrend);

    \draw[->,dashed]
        (\xpone+.1, \arrend-\arrlen)
        -- (\xpone+.1, \arrend);
    \draw[->,dashed]
        (\xpthree+.1, \arrend-\arrlen)
        -- (\xpthree+.1, \arrend);

    \draw[|-|] (\xpone-.1, \arrend-\arrlen-.3)
        --
        (\xptwo-.1, \arrend-\arrlen-.3)
        node[midway,fill=white,inner sep=0]
        {$\sigma\omega_1^{-1}$};
    \draw[|-|,dashed]
        (\xpone+.1, \arrend-\arrlen-.7)
        --
        (\xpthree+.1, \arrend-\arrlen-.7)
        node[midway,fill=white,inner sep=0]
        {$\sigma\omega_2^{-1}$};

    \node[draw,anchor=south west] (VAM1)
        at (\xptwo-.1, \yzero+\cabeza+.3)
        {VAM};
    \draw[->] (\xptwo-.1, \yzero+\cabeza)
        -- (VAM1.south west);
    \node[draw,anchor=south west] (VAM2)
        at (\xpthree-.1, \yzero+\cabeza+.3)
        {VAM};
    \draw[->] (\xpthree-.1, \yzero+\cabeza)
        -- (VAM2.south west);
    \node[draw,anchor=south west,dashed,
        fill=gray!20!white] (VAM2p)
        at (\xpthree+.1, \yzero+\cabeza+.25)
        {VAM};
    \draw[->,dashed]
        (\xpthree+.1, \yzero+\cabeza)
        -- (VAM2p.south west);
\end{tikzpicture}
        \caption{VAMs are sent when pedestrians
        move $\Delta$ meters
        at speed $\sigma$.}
        \label{fig:vamsdelta}
    \end{subfigure}
    \hfill
    \begin{subfigure}[b]{0.28\textwidth}
        \centering
        \begin{tikzpicture}
    \def\D{1.5}
    \def\xzero{.4}
    \def\yzero{1}

    \draw[|->] (0,0) -- (\xzero+2.5*\D,0)
        node[anchor=west] {$t$};

    \def\xpone{\xzero}
    \def\xptwo{\xzero+\D}
    \def\xpthree{\xzero+2*\D}
    \def\xpfour{\xzero+2*\D}
    \def\cabeza{.15}

    \pelao{\xpone}{\yzero}{\cabeza}
    \runpelao{\xptwo}{\yzero}{\cabeza}{-45}
    \runpelao{\xpthree}{\yzero}{\cabeza}{-22.5}

    \def\arrlen{.35}
    \def\arrend{-.1}

    \draw[->] (\xpone-.1, \arrend-\arrlen)
        -- (\xpone-.1, \arrend);
    \draw[->] (\xptwo-.1, \arrend-\arrlen)
        -- (\xptwo-.1, \arrend);
    \draw[->] (\xpthree-.1, \arrend-\arrlen)
        -- (\xpthree-.1, \arrend);

    \draw[->,dashed]
        (\xpone+.1, \arrend-\arrlen)
        -- (\xpone+.1, \arrend);
    \draw[->,dashed]
        (\xpthree+.1, \arrend-\arrlen)
        -- (\xpthree+.1, \arrend);

    \draw[|-|] (\xpone-.1, \arrend-\arrlen-.3)
        --
        (\xptwo-.1, \arrend-\arrlen-.3)
        node[midway,fill=white,inner sep=0]
        {$\omega_1^{-1}$};
    \draw[|-|,dashed]
        (\xpone+.1, \arrend-\arrlen-.7)
        --
        (\xpthree+.1, \arrend-\arrlen-.7)
        node[midway,fill=white,inner sep=0]
        {$\omega_2^{-1}$};

    \node[align=center,above] at (\xpone, 0)
        {$\sigma_1=.5$};
    \node[align=center,above] at (\xptwo, 0)
        {$\sigma_2=2$};
    \node[align=center,above] at (\xpthree, 0)
        {$\sigma_3=1.34$};

    \node[draw,anchor=south west] (VAM1)
        at (\xptwo-.1, \yzero+\cabeza+.3)
        {VAM};
    \draw[->] (\xptwo-.1, \yzero+\cabeza)
        -- (VAM1.south west);
    \node[draw,anchor=south west] (VAM2)
        at (\xpthree-.1, \yzero+\cabeza+.3)
        {VAM};
    \draw[->] (\xpthree-.1, \yzero+\cabeza)
        -- (VAM2.south west);
    \node[draw,anchor=south west,dashed,
        fill=gray!20!white] (VAM2p)
        at (\xpthree+.1, \yzero+\cabeza+.25)
        {VAM};
    \draw[->,dashed]
        (\xpthree+.1, \yzero+\cabeza)
        -- (VAM2p.south west);
\end{tikzpicture}
        \caption{VAMs are sent when
        pedestrians change their speed
        $\sigma$ (in \textrm{m/sec}).}
        \label{fig:vamssigma}
    \end{subfigure}
    \hfill
    \begin{subfigure}[b]{0.4\textwidth}
        \centering
        \begin{tikzpicture}
    \def\D{1.5}
    \def\xzero{.4}
    \def\yzero{.7}

    \draw[->] (0,0) -- (\xzero+3.5*\D,0)
        node[anchor=west] {$x$};
    \draw[->] (0,0) -- (0, 1)
        node[anchor=east] {$y$};

    \def\xpone{\xzero}
    \def\xptwo{\xzero+\D}
    \def\xpthree{\xzero+2*\D}
    \def\xpfour{\xzero+3*\D}
    \def\cabeza{.15}

    \pelao{\xpone}{\yzero}{\cabeza}
    \pelao{\xptwo}{\yzero}{\cabeza}
    \pelao{\xpthree}{\yzero+.5}{\cabeza}
    \pelao{\xpfour}{\yzero}{\cabeza}

    \draw[->] (\xpone+.3,.5*\yzero)
        -- (\xptwo-.3,.5*\yzero)
        node[midway,above] {$\theta_1$};
    \draw[->] (\xptwo+.3,.5*\yzero)
        -- (\xpthree-.3,1.25*\yzero)
        node[pos=.85,below] {$\theta_2$};
    \draw[thin,gray] (\xptwo+.3,.5*\yzero)
        -- (\xpthree,.5*\yzero);
    \draw[gray] (\xptwo+.75,.5*\yzero)
        arc (0:15:1);
    \draw[->] (\xpthree+.3,.1+\yzero)
        -- (\xpfour-.3,.5*\yzero)
        node[pos=.85,above] {$\theta_3$};
    \draw[thin,gray] (\xpthree+.3,.1+\yzero)
        -- (\xpthree+.9,.1+\yzero);
    \draw[gray] (\xpthree+.75,.1+\yzero)
        arc (0:-13:1);

    \def\arrlen{.35}
    \def\arrend{-.1}

    \draw[->] (\xpone-.1, \arrend-\arrlen)
        -- (\xpone-.1, \arrend);
    \draw[->] (\xptwo-.1, \arrend-\arrlen)
        -- (\xptwo-.1, \arrend);
    \draw[->] (\xpthree-.1, \arrend-\arrlen)
        -- (\xpthree-.1, \arrend);
    \draw[->] (\xpfour-.1, \arrend-\arrlen)
        -- (\xpfour-.1, \arrend);

    \draw[->,dashed]
        (\xpone+.1, \arrend-\arrlen)
        -- (\xpone+.1, \arrend);
    \draw[->,dashed]
        (\xpthree+.1, \arrend-\arrlen)
        -- (\xpthree+.1, \arrend);

    \draw[|-|] (\xpone-.1, \arrend-\arrlen-.3)
        --
        (\xptwo-.1, \arrend-\arrlen-.3)
        node[midway,fill=white,inner sep=0]
        {$\sigma\omega_1^{-1}$};
    \draw[|-|,dashed]
        (\xpone+.1, \arrend-\arrlen-.7)
        --
        (\xpthree+.1, \arrend-\arrlen-.7)
        node[midway,fill=white,inner sep=0]
        {$\sigma\omega_2^{-1}$};

    \node[draw,anchor=south west] (VAM1)
        at (\xptwo-.1, \yzero+\cabeza+.3)
        {VAM};
    \draw[->] (\xptwo-.1, \yzero+\cabeza)
        -- (VAM1.south west);
    \node[draw,anchor=south west] (VAM2)
        at (\xpthree-.1, .5+\yzero+\cabeza+.3)
        {VAM};
    \draw[->] (\xpthree-.1, .5+\yzero+\cabeza)
        -- (VAM2.south west);
    \node[draw,anchor=south west,dashed,
        fill=gray!20!white] (VAM2p)
        at (\xpthree+.1, .5+\yzero+\cabeza+.25)
        {VAM};
    \draw[->,dashed]
        (\xpthree+.1, .5+\yzero+\cabeza)
        -- (VAM2p.south west);
    \node[draw,anchor=south west] (VAM3)
        at (\xpfour-.1, \yzero+\cabeza+.3)
        {VAM};
    \draw[->] (\xpfour-.1, \yzero+\cabeza)
        -- (VAM3.south west);
\end{tikzpicture}
        \caption{VAMs are sent when
        pedestrians change their orientation
        $\theta$. Pedestrians move at speed
        $\sigma$.}
        \label{fig:vamstheta}
    \end{subfigure}
    \caption{VAMs generated by a pedestrian
    moving to the right.
    At the bottom we illustrate the positioning
    checks (arrows) at high frequency
    $\omega_1$\,\textrm{Hz} (solid)
    and low frequency
    $\omega_2$\,\textrm{Hz} (dashed).
    On top we illustrate the VAMs generated
    due to high frequency (solid)
    and low frequency checks (dashed).}
    \label{fig:vams}
\end{figure*}
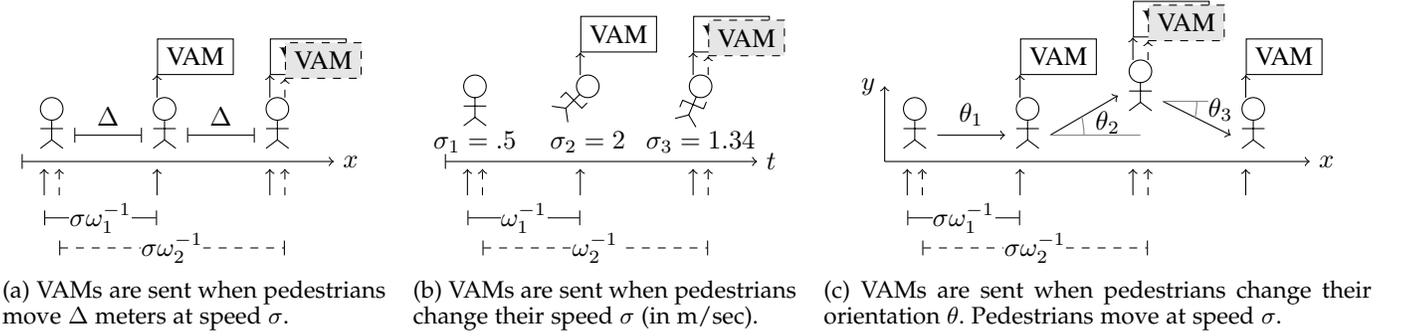


In this work we argue
whether the positioning sampling rate proposed
by ETSI (10\,\textrm{Hz})~\cite{etsi-va} is optimal to ensure pedestrian VAMs are
delivered upon channel congestion.
Namely, we investigate whether it is even possible
to decrease the positioning sampling
rate to generate less VAMs and
prevent congestion,
yet minimizing the time elapsed
between received pedestrian VAMs. Our work
sets forth the following contributions:
\begin{itemize}
    \item In \S\ref{sec:model}--\ref{sec:orient},
        we characterize how the positioning
        sampling rate impacts the rate
        at which pedestrians generate
        ETSI VAMs due to:
        position changes (\S\ref{sec:model}),
        speed changes
        (\S\ref{sec:speed-orient}), and
        heading/orientation changes
        (\S\ref{sec:orient}).
    \item In \S\ref{sec:minimize}, we
        formulate an optimization problem
        to find the positioning sampling
        rate minimizing the time elapsed
        between pedestrian VAM receptions. Our
        formulation accounts for the
        channel congestion. Moreover,
        we leverage
        our VAM characterization and propose
        an algorithm to find the
        optimal positioning sampling rate
        for 802.11p  (\S\ref{subsec:80211p}).
    \item In \S\ref{sec:results},
        we ($i$)
        show that our VAM characterization
        is conservative with an error
        below $0.157$\,\textrm{VAMs/sec}
        (\S\ref{subsec:validation});
        and ($ii$) show through 802.11p
        numerical/simulation results
        how our algorithm minimizes the
        time between pedestrian VAMs 
        decreasing a 30\% ETSI
        positioning sampling rate
        (\S\ref{subsec:numerical} and
        \ref{subsec:simulation}).
\end{itemize}
The motivation behind using 802.11p is
three-folded: its integration in
cars (e.g., Volkswagen ID.4, and a fleet of Renault cars~\cite{renault}) and implementation
of commercial
Onboard Units~\cite{Commsignia}; 
the existence of a
closed-form analytical model of the
channel access~\cite{BaiocchiAoI}; and
the stability of the
Artery/Veins/SUMO~\cite{Artery,Veins,sumo2012}
stack to emulate ETSI VRUs and VAMs
on top of 802.11p.
Nevertheless, it is possible to adapt
our solution to technologies as
802.11bd or C-V2X --
see the end
of \S\ref{subsec:80211p} for further details.


\section{Overview of ETSI VAM generation rules}
\label{sec:vam_rules}

VAMs are generated depending on VRU kinematic triggers. The interval between two
VAMs (\textit{T\_GenVam})
is in between 0.1\,\textrm{sec}
(\textit{T\_GenVamMin}) and 5\,\textrm{sec}
(\textit{T\_GenVamMax})
-- see~\cite[Table~16]{etsi-va}. VRUs must check at least every \textit{T\_GenVamMin} seconds whether a VAMs should be
generated~\cite[6.2]{etsi-va}.
That is, VRUs check their position at a rate $\omega\geq1/\textit{T\_GenVamMin}$.

A VAM is generated by either of two conditions:
\begin{enumerate}
    \item the VRU exceeds a set of kinematic thresholds with respect to the
        last VAM:
        \begin{enumerate}[label=(\alph*)]
            \item its position ($x$)
                changed by more than
                \textit{minReferencePointPositionChangeThreshold} ($\Delta$) meters;\label{threshold:position}
            \item its speed ($\sigma$) changed 
                by more than
                \textit{minGroundSpeedChangeThreshold}
                ($\delta_\sigma$)
                \textrm{m/sec}; or\label{threshold:speed}
            \item its orientation ($\theta$)
                changed by more than
                \textit{minGroundVelocityOrientationChangeThreshold}
                ($\delta_\theta$) degrees; or\label{threshold:orient}
        \end{enumerate}
    \item the time since the last VAM exceeds \textit{T\_GenVamMax} and rules for redundance mitigation are met (e.g., if the VRU is not stationary, up to 10 VAMs can be skipped).
\end{enumerate}

Fig.~\ref{fig:vams} illustrates the VAMs
that pedestrians generate due to the
aforementioned kinematic thresholds
--- i.e. \ref{threshold:position},
\ref{threshold:speed}, and
\ref{threshold:orient}.
The values set by
ETSI~\cite[Table~17]{etsi-va} are
$\Delta=4$\,\textrm{meters},
$\delta_\sigma=0.5$\,\textrm{m/sec} and
$\theta=4$\,\textrm{degrees}.

\begin{table}[t]
    \centering
    \caption{Notation Table}
    \begin{tabular}{c l}
        \toprule
        \textbf{Symbol} & \textbf{Definition}\\
        \midrule
        $\omega$ & Positioning sampling frequency\\
        $x(t)$ & Pedestrian position at time $t$\\
        $\sigma(t)$ & Pedestrian speed at time $t$\\
        $\theta(t)$ & Pedestrian orientation at time $t$\\
        $\Delta$ & Position difference threshold\\
        $\delta_\sigma$ & Speed difference threshold\\
        $\delta_\theta$ & Orientation difference threshold\\
        $\lambda_\Delta(\omega)$ & Rate of position change VAMs with sampling $\omega$\\
        $\lambda_\sigma(\omega)$ & Rate of speed change VAMs with sampling $\omega$\\
        $\lambda_\theta(\omega)$ & Rate of orientation change VAMs with sampling $\omega$\\
        $\lambda(\omega)$ & Total VAM rate with sampling $\omega$\\
        \bottomrule
    \end{tabular}
\end{table}

A VRU will trigger more VAMs if it checks
more frequently its position, thus
the triggering conditions.
For example, the VRU generates more VAMs
with
$\text{\textit{T\_GenVamMin}}=100$\,\textrm{ms}
($\omega=10$\,\textrm{Hz})
than with
$\text{\textit{T\_GenVamMin}}=1$\,\textrm{sec}
($\omega=1$\,\textrm{Hz}).
This is because it is more likely
to envision changes of position,
speed or orientation if
the position is sampled more often.
In other words, the VAM rate of a pedestrian
$\lambda(\omega)$ is monotonically
increasing on the checking rate --- i.e.
$\lambda(\omega_1)\leq\lambda(\omega_2)$
with $\omega_1\leq\omega_2$.

Having high VAM rates helps to have
more accurate and real time
information about where pedestrians are,
how fast they move and where they are heading
to. However, high VAM rates may result
in collisions in the wireless medium
when using technologies as 802.11p.
Therefore, in this article we look for
the best positioning sampling rate $\omega$ to
have accurate information about the
VRU, yet preventing the collision of the VAMs.

Before looking for the best
positioning sampling rate $\omega$,
in the following sections
we characterize the rate of VAMs
due to position changes in \S\ref{sec:model},
speed changes in \S\ref{sec:speed-orient} and
orientation changes in \S\ref{sec:orient}.

\section{VAMs upon position changes}
\label{sec:model}

As stated in \S\ref{sec:vam_rules},
VRUs check at a rate $\omega$
whether they have moved
$\Delta$ meters or not.
If the VRU travels
at speed $\sigma$\,\textrm{meters/sec},
each checking period it
travels
$\sigma\omega^{-1}$\,\textrm{meters}
-- as illustrated in~Fig.\ref{fig:vamsdelta}.
Moreover, it takes
$\lceil \omega\Delta/\sigma \rceil$
positioning samples to realize that a
pedestrian exceeded
$\Delta$\,\textrm{meters}.
As a result, the rate at which a pedestrian
generates VAMs due to position changes is
\begin{equation}
    \lambda_\Delta(\omega)=\omega
    \left\lceil
        \frac{\omega\Delta}{%
        \sigma}
    \right\rceil^{-1}
    \label{eq:lambda}
\end{equation}

\begin{figure}[t]
    \centering
    \begin{tikzpicture}
\begin{axis}[
    domain=0:15,
    samples=1000, 
    axis lines = left,
    xlabel = {$\omega$ [Hz]},
    ylabel = {$\lambda_\Delta$ [VAM/sec]},
    grid,
    height=.6\columnwidth,
    width=\columnwidth,
    legend style={at={(0.07,1.05)},anchor=south west, legend columns=2,draw=black,fill=none}
]

\path[fill=gray!20]
    (axis cs:10,0) -- (axis cs:10,2)
    -- (axis cs:15,2) -- (axis cs:15,0);
\draw[dashed] (axis cs:10,0) -- (axis cs:10,2)
    node[pos=.45,align=center,anchor=west]
    {ETSI~\cite[6.2]{etsi-va}\\positioning\\sampling rates};

\addplot [
    Firebrick2,
    very thick
]
{x / ceil(4 * x / 1.34)};
\addlegendentry{$\Delta=4,\sigma=1.34$}

\addplot [
    DodgerBlue2,
    very thick
]
{x / ceil(4 * x / .67)};
\addlegendentry{$\Delta=4,\sigma=0.67$}

\addplot [
    Green3,
    very thick
]
{x / ceil(2 * x / 1.34)};
\addlegendentry{$\Delta=2,\sigma=1.34$}

\addplot [
    Gold2,
    very thick
]
{x / ceil(2 * x / 2.68)};
\addlegendentry{$\Delta=2,\sigma=2.68$}

\end{axis}
\end{tikzpicture}
    \vspace{-1em}
    \caption{VAM generation
    rate due to position changes
    $\lambda_\Delta$ increases
    in a saw teeth fashion
    with the positioning sampling rate
    $\omega$. The
    smaller the distance
    $\Delta$ and the
    larger the speed $\sigma$,
    the larger $\lambda_\Delta$.
    We assume a constant speed $\sigma$.}
    \label{fig:saw}
\end{figure}
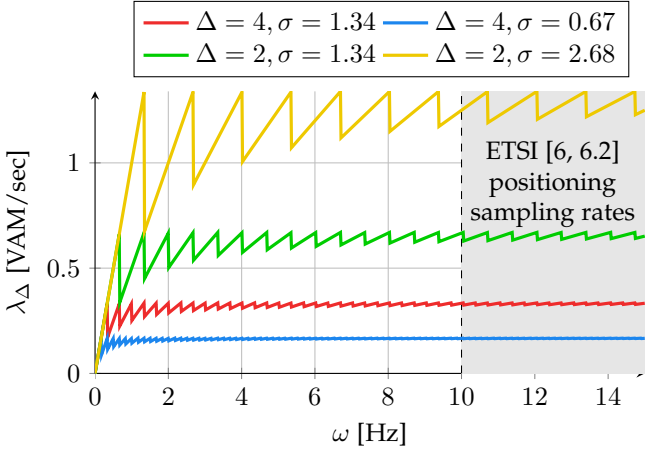
The VAM generation rate
$\lambda_\Delta$
upon position changes
exhibits a saw-teeth
behaviour that depends on
the checking distance
$\Delta$, speed $\sigma$
and checking frequency
$\omega$.
Fig.~\ref{fig:saw} illustrates
the saw-teeth tendency and
highlights how position VAMs
are generated more frequently
if the average speed increases.
That is, it is more likely to
realize a change of $\Delta$\,\textrm{meters}
in the position if the pedestrian
moves faster.

Note that in~\eqref{eq:lambda} we assume
constant speeds for pedestrians.
The assumption is supported by the
rather negligible speed variations
that we foresee in \S\ref{sec:speed-orient}
analysis. Despite the assumption,
Fig.~\ref{fig:saw} evidences
that the rate of positioning VAMs
$\lambda_\Delta$ stabilizes
at rather small values of $\omega$
-- e.g. around $\omega=2$ for
average pedestrian speeds~\cite{avgspeed} of
$\sigma=1.34$\,\textrm{m/sec} and
the standard~\cite{etsi-va}
$\Delta=4$\,\textrm{meters}.
Hence, upon rather constant speeds
pedestrians can decrease the rate at
which they check the positioning without
harnessing the generation of VAMs upon
changes of positions. In other
words, as Fig.~\ref{fig:saw}
illustrates, it is not necessary to set
$\omega\geq10$\,\textrm{Hz} --
as specified by ETSI in~\cite[6.2]{etsi-va}.

\section{VAMs upon speed changes}
\label{sec:speed-orient}

In the previous section we
consider that VRUs travel at a constant
speed $\sigma$\,\textrm{m/sec}
to derive the rate of VAMs
generated because the VRU
position changes by more than
$\Delta$\,\textrm{meters}.
In real-world the VRUs vary
their speed and we should
consider a
time dependent
variable $\sigma(t)$.

A speed change will trigger a VAM if
$|\sigma(i\omega^{-1})
-\sigma(i_0\omega^{-1})|
\geq\delta_\sigma$ holds
for $i>i_0\geq0$ --
with $\delta_\sigma$ the
\emph{minGroundSpeedChangeThreshold}
defined in~\cite{etsi-va}.
That is, a VAM is generated if at
the $i$\textsuperscript{th} sample
the speed changed by more than
$\delta_\sigma=0.5$\,\textrm{m/sec}
\cite{etsi-va}
with respect to the speed at the
$i_0$\textsuperscript{th} sample.


We resort to the Gradient
Navigation Model
(GNM)~\cite{pedestrians}
to understand how pedestrians
move and change their speed
$\sigma(t)$ over time.
According to the GNM, every
pedestrian movement is
affected by the neighboring
pedestrians, e.g. a pedestrian
will not move to the right
if there is another person
at its near right; otherwise
it would clash with that
person. 
We refer to
$N(t)\in\mathbb{R}^2$ as the
desired direction of
a pedestrian given its
neighbors movement and the
obstacles around it, which
is defined
in~\cite{pedestrians} as
\begin{equation}
    N(t)=g\left(
        g(N_T)
        +g(N_P)
    \right)
    \label{eq:desired}
\end{equation}
with
$g:\mathbb{R}^2\mapsto
\mathbb{R}^2$,
$\|g(x)\|\leq1,\forall x$,
$N_T$ the target vector
(where the pedestrian goes),
and $N_P$ the perturbance
of neighboring pedestrians
-- see~\cite[(2),(A4)]{pedestrians}
for the precise definitions
of $N_P$ and $g(x)$,
respectively

According to~\cite{pedestrians},
the speed vector at a given time $\dot{x}(t)$
is given by the product of the target vector
-- 
with norm~\eqref{eq:desired},
$||N(t)||\leq1$
--
and the relaxed speed
$w(t)$ that the pedestrian
has. Hence,
the Ordinary Differential
Equations (ODE) system
is
\begin{align}
    \dot{x}(t) &=
    w(t)N(t)\label{eq:x-der}\\
    \dot{w}(t) &=
    \frac{1}{\tau}
    \left(
    v||N(t)||-w(t)
    \right)\label{eq:w-der}
\end{align}
with $v$ the desired speed, and
$\tau$ the speed of
reaction of a pedestrian,
i.e. how long it takes
to adapt her speed.

Note that the ODE system
in~\eqref{eq:x-der}-\eqref{eq:w-der}
fully determines what is
$\sigma(i\omega^{-1})
=\|\dot{x}(i\omega^{-1})\|$
and it is enough to resort
to approaches as the Euler
method or
Dormand-Prince-45~\cite{dormand}
to find
$\sigma(i\omega^{-1})$.

Given a reference sample when the
last VAM was generated $i_0$,
it is possible to know when the
speed will change by more than
$\delta_\sigma$\,\textrm{m/sec}.
In particular,
we obtain the time to the next
VAM due to a speed change as
$\omega^{-1}
\inf_{i>i_0}
\left\{
    \left|
    \sigma(i\omega^{-1})
    - 
    \sigma(i_0\omega^{-1})
    \right|
    \geq\delta_\sigma
\right\}$. That is, we evolve
the ODE system until the first
sample $i>i_0$ at which the speed
change exceeds $\delta_\sigma$.
Consequently, we estimate the rate
of VAMs due to speed changes as the
inverse of the time until the next
VAM occurs given the reference sample
$i_0$

\begin{lemma}[VAM rate due
to speed changes]
    A pedestrian with
    initial speed $w(0)$,
    reaction time $\tau$,
    sampling rate $\omega$
    and desired speed $v$
    generates VAMs due to
    speed changes at rate
    \begin{multline}
        \lambda_\sigma
        (\omega)=
        \omega\bigg(
        \inf_{i>i_0}
        \Big\{
        \big|
        N_i a^{\frac{i}{h\omega}}
        w(0) \\
        +v\frac{1-a}{a}
        N_i\sum_{m=0}^{\frac{i}{h\omega}}
        a^{\frac{i}{h\omega}-m}
        N_{mh}
        - w(i_0\omega)
        N_{i_0}
        \big|
        \geq \delta_\sigma
        \big\}
        \bigg)^{-1}
        \label{eq:lambda-speed-lemma}
    \end{multline}
    with
    $a=1-\tfrac{h}{\tau},
    N_i=\|N(i h)\|$ and
    $h<\tau$
    the Euler method
    step size.
    \label{lemma:speed}
\end{lemma}

\begin{proof}
    Using the Euler method
    in the ODE
    system~\eqref{eq:x-der}-\eqref{eq:w-der}
    we obtain $w(t)$ by
    induction.

    Considering the step
    size $h>0$ we have
    a one step progression
    \begin{equation}
        w(h)=
        w_0\left(
            1-\frac{h}{\tau}
        \right)
        +\frac{hv}{\tau}
        \|N(0)\|
        =w(0)a+v(1-a)
        N_0
    \end{equation}

    If we take another step
    it is easy to show that
    \begin{equation}
        w(2h)
        =a^2 w(0)
        + v(1-a)\left(
            aN_0+N_h
        \right)
    \end{equation}
    so we obtain
    $w(kh)$ by induction on
    $k$
    \begin{equation}
        w(kh)
        =a^{k}w(0)
        +v\frac{1-a}{a}
        \sum_{m=0}^k
        a^{k-m} N_{m h}
        \label{eq:kth-term}
    \end{equation}

    By doing a change of
    variable $t=kh$ and
    then taking
    $t=i\omega^{-1}$
    in~\eqref{eq:kth-term}
    we get
    \begin{equation}
        w(i\omega^{-1})
        =a^{\frac{i}{h\omega}}
        w(0)+v\frac{1-a}{a}
        \sum_{m=0}^{\frac{i}{h\omega}}
        a^{\frac{i}{h\omega}-m}
        N_{m h}
    \end{equation}
    and~\eqref{eq:lambda-speed-lemma}
    holds since
    $\sigma(i\omega^{-1})
        =w(i\omega^{-1})
        N_{i}$
\end{proof}

\begin{figure*}[t]
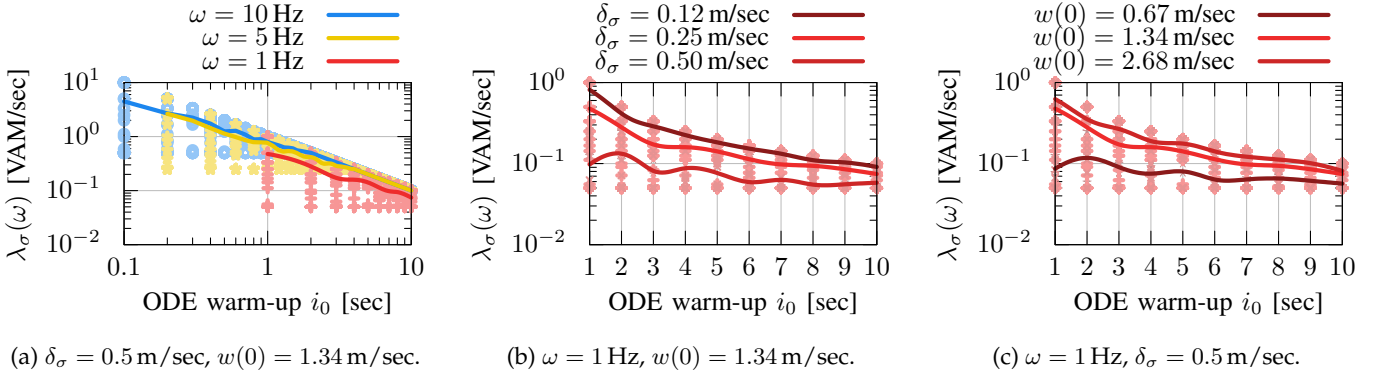

   \begin{subfigure}[b]{0.32\textwidth}
       \centering
       \input{fig/lambda-omega}
       \vspace{-1em}
       \caption{$\delta_\sigma=0.5$\,\textrm{m/sec},
       $w(0)=1.34$\,\textrm{m/sec}.}
       \label{fig:lambda-omega}
   \end{subfigure}
   \hfill
   \begin{subfigure}[b]{0.32\textwidth}
       \centering
       \input{fig/lambda-speed}
       \vspace{-1em}
       \caption{$\omega=1$\,\textrm{Hz},
       $w(0)=1.34$\,\textrm{m/sec}.}
       \label{fig:lambda-speed}
   \end{subfigure}
   \hfill
   \begin{subfigure}[b]{0.32\textwidth}
       \centering
       \input{fig/lambda-wi0}
       \vspace{-1em}
       \caption{$\omega=1$\,\textrm{Hz},
       $\delta_\sigma=0.5$\,\textrm{m/sec}.}
       \label{fig:lambda-wi0}
   \end{subfigure}
   \caption{
    Speed VAM generation
    rate $\lambda_\sigma(
    \omega)$ vs. the
    pedestrian
    GNM~\cite{pedestrians}
    warm-up time $i_0$.
   Pedestrians have
   reaction time
   $\tau=0.5$\,\textrm{sec}
   and desired speed
   $v=1.34$\,\textrm{m/sec}.
   We try out different:
   (a) 
   checking rates $\omega$;
   (b) {\em minGroundSpeedChangeThreshold}
   $\delta_\sigma$; and
   (c) initial speeds
   $w(0)$. We illustrate the VAM rate
   for each random trajectory of the ODE
   system (points), and the average VAM
   rate (lines).
   }
   \label{fig:ode}
\end{figure*}

Although it may seem like
Lemma~\ref{lemma:speed}
provides an intractable expression
for $\lambda_\sigma(\omega)$, it
allows us to understand
the dynamics of a pedestrian.
In particular, we observe
how it becomes more difficult
to exceed the
{\em minGroundSpeedChangeThreshold}
parameter $\delta_\sigma$
as we increase $i_0$
(the ODE warm-up time).
This is mainly because
the term $a$ is below one
and it is powered to
$i>i_0$. Hence, the
speed differences reduce
as time passes.

Moreover, given that $\|N(t)\|\leq1$
it is possible to randomize the
target vector and estimate
$\lambda_\sigma(\omega)$.
Specifically, we
i) take random realizations of
$N_{i_0},N_{i_0+1},\ldots$; and
ii) compute $\lambda_\sigma(\omega)$
for each random realization.
Each random realization gives 
a possible trajectory of he pedestrian
corresponding to a set of initial
conditions. Thus, we resort to the
average of such random realizations
to estimate $\lambda_\sigma(\omega)$.

In Fig.~\ref{fig:ode} we illustrate
how the speed VAMs rate estimation
changes with respect to the
ODE warmup $i_0$, sampling rate $\omega$,
$\delta_\sigma$, and initial speed
$w(0)$. Results show the aforementioned
phenomena of having smaller rate as
the ODE warmup increases. On top, results
also evidence that the higher the sampling
rate $\omega$, the higher the VAM
rate due to speed changes
$\lambda_\sigma(\omega)$ because more
checks can detect more speed changes.

Later, in \S\ref{subsec:validation},
we show that the randomized approach
to estimate $\lambda_\sigma(\omega)$
results into upper bounds for the
rate of VAMs generated in a validated
mobility simulator implementing the
GNM~\cite{pedestrians}. Such upper
bounds hold for rates above
$\omega=10^{-1}$\,\textrm{Hz}, thus
covering the range of sampling rates
considered
by ETSI~\cite{etsi-va}
$\omega\geq10$\,\textrm{Hz}.




\section{VAMs upon orientation changes}
\label{sec:orient}

In this section, we want to
know the rate at which VAMs
are generated due to
orientation changes. That is,
the VAMs generated when
the user orientation
$\theta(t)$ changed more than
$\delta_\theta=4º$~\cite{etsi-va}
with respect to the orientation
reported in the last VAM at time
$\theta(i_0\omega^{-1})$.


Without loss of generality
we take as reference the
vector $(1,0)$, hence
define the orientation at
time $t=i\omega^{-1}$ as
\begin{equation}
    \theta(i\omega^{-1})
    =\arccos\left(\frac{%
    \left(x(i\omega^{-1})-
    x((i-1)\omega^{-1})
    \right)\cdot
    (1,0)
    }{\left|
    x(i\omega^{-1})-
    x((i-1)\omega^{-1})
    \right|}
    \right)
    \label{eq:orient}
\end{equation}
To obtain what will be the
orientation~\eqref{eq:orient}
at time $i\omega^{-1}$ we resort to the
GNM~\cite{pedestrians} ODE system
in~\eqref{eq:x-der}-\eqref{eq:w-der}.

With the orientation
$\theta(i\omega^{-1}),\ i>0$ we then
obtain the rate of VAMs due to orientation
changes $\lambda_\theta$. In particular,
in \S\ref{subsec:neigh-influence} we explain
how pedestrians and obstacles influence
the pedestrian position. Then, in
 \S\ref{subsec:ppp} we obtain the average
change of orientation that a pedestrian
experiments when neighbors coordinates
follow a Poisson Point Process (PPP).
Lastly, in \S\ref{subsec:avg-rate} we use
the average change of orientation to
derive the rate of VAMs due to speed
changes $\lambda_\theta$.

\subsection{Influence of Neighboring
Pedestrians}
\label{subsec:neigh-influence}

As shown in~\eqref{eq:orient}, the
change of orientation is computed taking
the difference between the current
position $x(i\omega^{-1})$ and
the prior position $x((i-1)\omega^{-1})$.
Based on the GNM~\eqref{eq:x-der}, the
direction of the position vector $x(t)$
is fully determined by the pedestrian
desired direction $N(t)$. 

The pedestrian desired direction $N(t)$
is impacted~\eqref{eq:desired}
by repulsive forces induced by neighboring
pedestrians $\nabla P_i$ and obstacles
$\nabla P_b$. The negative sum of such
repulsions results in the repulsive forve
$N_P(t)=g(-\sum_k\nabla P_k)$
that alters the pedestrian desired
direction
-- see~Fig.~\ref{fig:repulsion}.

According to~\cite{pedestrians},
the repulsion induced by
a neighboring pedestrian $i$ is
\begin{equation}
    \nabla P_i
    = h\left(
        \|x_i\|; p_i,R_i
    \right)
    s(\theta_i)
    \frac{x_i}{\|x_i\|}
    \label{eq:repulsion}
\end{equation}
with $h(\cdot)$ a
monotonically decreasing
function w.r.t. the distance
taking its maximum at
$p_i$ and zero for values
above $R_i$
-- see~\cite[(3)]{pedestrians}--
and $s(\theta_i)$ the
repulsion of neighbors
depending on the angle
angle $\theta_i$
w.r.t. pedestrian $i$
-- see~\cite[(4)]{pedestrians}.

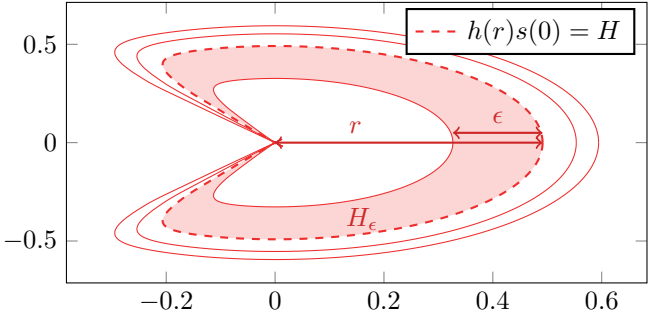
\begin{figure}[t]
   \centering
    \begin{tikzpicture}

\pgfmathsetmacro{\pi}{3.59}; 
\pgfmathsetmacro{\Ri}{0.7}; 
\pgfmathsetmacro{\Rdi}{0.03}; 
\pgfmathsetmacro{\ki}{0.6}; 
\pgfmathsetmacro{\xoi}{0.3}; 

\pgfplotsset{
    compat=1.12,
    /pgf/declare function={
        l(\phi,\h,\p,\R,\Rd,\k,\xo) = %
            1/(ln(\h/\p*(1+exp(-(cos(\k*\phi)-\xo)/\Rd))));
    },
    /pgf/declare function={
        f(\phi,\h,\p,\R,\Rd,\k,\xo) = %
            l(\phi,\h,\p,\R,\Rd,\k,\xo) > -1 ?%
                (l(\phi,\h,\p,\R,\Rd,\k,\xo) < 0 ? %
                    sqrt( (\R)^2* ( l(\phi,\h,\p,\R,\Rd,\k,\xo)+1  ) ) %
                        : 0) : 0  ;
    },
}

    \begin{axis}[thick,smooth,no markers,
        trig format plots=rad,
        width=1.05\columnwidth,
        height=.6\columnwidth,
    ]

        \foreach \hh in {.1, .25} {

            \addplot[name path=A,data cs=polar,Firebrick2,domain=0:2*pi,
                samples=360,smooth,
                forget plot
            ]
                (\x,{ f(\x,\hh,\pi,\Ri,\Rdi,\ki,\xoi) })
                coordinate[pos=0] (As)
                coordinate[pos=.4] (Bs);
            \addplot[name path=B,data cs=polar,Firebrick2,domain=0:-2*pi,
                samples=360,smooth,
                forget plot
            ]
                (\x,{ f(\x,\hh,\pi,\Ri,\Rdi,\ki,\xoi) });

        }

        \addplot[name path=A,data cs=polar,Firebrick2,domain=0:2*pi,
            samples=360,smooth,dashed,thick,
            line legend
        ]
            (\x,{ f(\x,.5,\pi,\Ri,\Rdi,\ki,\xoi) })
            coordinate[pos=0] (As)
            coordinate[pos=.4] (Bs);
        \addplot[name path=B,data cs=polar,Firebrick2,domain=0:-2*pi,
            samples=360,smooth,dashed,thick,
            forget plot
        ]
            (\x,{ f(\x,.5,\pi,\Ri,\Rdi,\ki,\xoi) });

        \addplot[name path=A2, data cs=polar,Firebrick2,domain=0:2*pi,
            samples=360,smooth,
            forget plot
        ]
            (\x,{ f(\x,1,\pi,\Ri,\Rdi,\ki,\xoi) })
            coordinate[pos=0] (As2);
        \addplot[name path=B2, data cs=polar,Firebrick2,domain=0:-2*pi,
            samples=360,smooth,
            forget plot
        ]
            (\x,{ f(\x,1,\pi,\Ri,\Rdi,\ki,\xoi) });
        \addplot[data cs=polar,transparent,domain=0:-2*pi,
            samples=360,smooth,
            forget plot
        ]
            (\x,{ f(\x,.75,\pi,\Ri,\Rdi,\ki,\xoi) })
            coordinate[pos=.4] (regNode);

        \addplot[Firebrick2!20, forget plot] fill between[of=A and A2];
        \addplot[Firebrick2!20, forget plot] fill between[of=B and B2];

        \draw[<->,Firebrick3,thick] (0,0) -- (As)
            node[pos=.3,above] {$r$};

        \coordinate (As2top) at
            ($(As2) + (axis direction cs:0,.05)$);
        \coordinate (Astop) at
            ($(As) + (axis direction cs:0,.05)$);

        \draw[<->,Firebrick3,thick]
            (As2top) -- (Astop)
            node[pos=.5,above] {$\epsilon$};

        \node[Firebrick3] at (regNode) {$H_\epsilon$};

        \legend{$h(r)s(0)=H$};
    \end{axis}

\end{tikzpicture}
    \caption{Isolines of
    influence $h(r)s(\theta)$
    for a pedestrian at
    the origin moving to
    the right.
    $H_\epsilon$ refers
    to the band of influence
    of width $\epsilon$
    w.r.t. isoline of
    magnitude $H$.}
    \label{fig:isolines}
\end{figure}

Fig.~\ref{fig:isolines}
shows the isolines of
influence~\eqref{eq:repulsion}
experienced by a pedestrian
at the origin. The isolines
$h(\cdot)s(\cdot)=H$
evidence
a cardioid shape that capture how a 
pedestrian does not experience repulsion
$\nabla P_i=0$ if the pedestrian $i$
is behind her.

Note that in~\eqref{eq:repulsion}
we consider that the
pedestrian of interest is
located at the origin
$x_0=(0,0)$ and the original
expression for
$h(\|x_i-x_0\|;p_i,R_i)$
in~\cite[(5)]{pedestrians}
is simplified
to~\eqref{eq:repulsion}.
Additionally, we also
use~\eqref{eq:repulsion}
to define the repulsion
$\nabla P_b$
induced by an obstacle $b$
as the wall in
Fig.~\ref{fig:repulsion},
with the peculiarity of
replacing $p_i,R_i$
by $p_b,R_b$ -- i.e. the
maximum value and support
of the function $h(\cdot)$.

Overall, we compute the desired
direction at a given time
$N(t)$ in~\eqref{eq:orient}
using the pedestrian
direction $N_T$, neighbors/obstacles
repulsion
$N_P=-\sum_b\nabla P_b
-\sum_i \nabla P_i$
and the normalizing
function
$g:\mathbb{R}^2\to
\mathbb{R}^2$
with $\|g(x)\|\in[0,1]$
-- see its definition
in~\cite[Appendix~A]{pedestrians}.
Consequently, we can
obtain the change of
orientation by
replacing $x(t)$ with
$N(t)$ in~\eqref{eq:orient},
since the pedestrian
position $x(t)$ direction
is fully determined by
$N(t)$
-- see~\eqref{eq:x-der}.

In Appendix~\ref{app:bound} we try to
bound the orientation $\theta(i\omega^{-1})$
hoping that the orientation of a user
stabilizes as time evolves, hence resulting
into $\lambda_\theta\simeq0$. For we could not
find any bound, in \S\ref{subsec:ppp} we use the pedestrians repulsion
$\nabla P_i$ to compute the
average change of orientation
$\mathbb{E}[\theta((i+k)\omega^{-1})
-\theta(i\omega^{-1})]$ and get
the rate of orientation change VAMs
$\lambda_\theta$.

\subsection{Average
Change of Orientation}
\label{subsec:ppp}

\begin{figure}[t]
   \centering
    \begin{tikzpicture}

\pgfmathsetmacro{\pi}{3.59}; 
\pgfmathsetmacro{\Ri}{0.7}; 
\pgfmathsetmacro{\Rdi}{0.03}; 
\pgfmathsetmacro{\ki}{0.6}; 
\pgfmathsetmacro{\xoi}{0.3}; 
\pgfmathsetmacro{\d}{0.5}; 

\pgfplotsset{
    compat=1.12,
    /pgf/declare function={
        hs(\phi,\dis,\p,\R,\Rd,\k,\xo) = %
        \p*exp(1/((\dis/\R)^2-1)) %
        * 1/(1+exp(-(cos(\k*\phi)-\xo)/\Rd));
    },
    /pgf/declare function={
        l(\phi,\h,\p,\R,\Rd,\k,\xo) = %
            1/(ln(\h/\p*(1+exp(-(cos(\k*\phi)-\xo)/\Rd))));
    },
    /pgf/declare function={
        f(\phi,\h,\p,\R,\Rd,\k,\xo) = %
            l(\phi,\h,\p,\R,\Rd,\k,\xo) > -1 ?%
                (l(\phi,\h,\p,\R,\Rd,\k,\xo) < 0 ? %
                    sqrt( (\R)^2* ( l(\phi,\h,\p,\R,\Rd,\k,\xo)+1  ) ) %
                        : 0) : 0  ;
    },
    /pgf/declare function={
        fdown(\phi,\h,\p,\R,\Rd,\k,\xo) = %
            min( abs(\d/sin(\phi)),  f(\phi,\h,\p,\R,\Rd,\k,\xo));
    },
    /pgf/declare function={
        norm(\x,\y)=sqrt((\x)^2 + (\y)^2);
    },
    /pgf/declare function={
        mol(\x,\R,\p)=max(0,%
            exp(1)*exp( 1/((\x/\R)^(2*\p)-1) ) );
    },
    /pgf/declare function={
        rmol(\x,\p)=mol(\x,1,\p)*\x + (1-mol(\x,1,\p));
    },
    /pgf/declare function={
        gscale(\x,\y,\p)=1/norm(\x,\y)*rmol(norm(\x,\y),\p);
    },
}

    \begin{axis}[thick,smooth,no markers,
        trig format plots=rad,
        width=1.05\columnwidth,
        height=.6\columnwidth,
        xmin=-.75,
        xmax=1,
        ymax=1.1,
    ]

        \foreach \hh in {.1, .25} {

            \addplot[name path=A,data cs=polar,Firebrick1!30,domain=0:2*pi,
                samples=360,smooth,
                forget plot
            ]
                (\x,{ f(\x,\hh,\pi,\Ri,\Rdi,\ki,\xoi) })
                coordinate[pos=0] (As)
                coordinate[pos=.4] (Bs);
            \addplot[name path=B,data cs=polar,Firebrick1!30,domain=0:-2*pi,
                samples=360,smooth,
                forget plot
            ]
                (\x,{ fdown(\x,\hh,\pi,\Ri,\Rdi,\ki,\xoi) });

        }

        \addplot[name path=A,data cs=polar,Firebrick1!30,domain=0:2*pi,
            samples=360,smooth,
            line legend
        ]
            (\x,{ f(\x,.5,\pi,\Ri,\Rdi,\ki,\xoi) })
            coordinate[pos=0] (As)
            coordinate[pos=.4] (Bs);
        \addplot[name path=B,data cs=polar,Firebrick1!30,domain=0:-2*pi,
            samples=360,smooth,
            forget plot
        ]
            (\x,{ fdown(\x,.5,\pi,\Ri,\Rdi,\ki,\xoi) });

        \addplot[name path=A2, data cs=polar,Firebrick1!30,domain=0:2*pi,
            samples=360,smooth,
            forget plot
        ]
            (\x,{ f(\x,1,\pi,\Ri,\Rdi,\ki,\xoi) })
            coordinate[pos=0] (As2);
        \addplot[name path=B2, data cs=polar,Firebrick1!30,domain=0:-2*pi,
            samples=360,smooth,
            forget plot
        ]
            (\x,{ fdown(\x,1,\pi,\Ri,\Rdi,\ki,\xoi) });
        \addplot[data cs=polar,transparent,domain=0:-2*pi,
            samples=360,smooth,
            forget plot
        ]
            (\x,{ fdown(\x,.75,\pi,\Ri,\Rdi,\ki,\xoi) })
            coordinate[pos=.4] (regNode);

        \addplot[name path=wallu,domain=-.6:.8,gray] {-\d}
            coordinate[pos=.5] (wt);
        \addplot[name path=walld,domain=-.6:.8,gray] {-(\d+.15)}
            coordinate[pos=.5] (wb);
        \addplot[fill=gray!20, forget plot,
            pattern=north east lines,
            pattern color=gray!40]
            fill between[of=wallu and walld];
        \draw[transparent] (wt) -- (wb)
            node[midway] (wallLab) {};
        \node[black!20!gray] at (wallLab) {wall};

        \pgfmathsetmacro{\nearpa}{-20};
        \pgfmathsetmacro{\nearph}{.5}; 
        \pgfmathsetmacro{\nearpr}{fdown(\nearpa,\nearph,\pi,\Ri,\Rdi,\ki,\xoi)}; 
        \pgfmathsetmacro{\nearpx}{\nearpr*cos(\nearpa)};
        \pgfmathsetmacro{\nearpy}{\nearpr*sin(\nearpa)};
        \pgfmathsetmacro{\nearpd}{sqrt((\nearpx)^2+(\nearpy)^2)};
        \pgfmathsetmacro{\nearrepx}{-\nearph/\nearpd*\nearpx};
        \pgfmathsetmacro{\nearrepy}{-\nearph/\nearpd*\nearpy};

        \coordinate (nearp) at (\nearpx, \nearpy);
        \node[circle,fill=Firebrick4,inner sep=0pt,
            minimum size=6pt, color=Firebrick2,
            label=west:{\color{Firebrick2}$x_i$}]
            at (nearp) {};

        \draw[thick,color=Firebrick2,
            line cap=round,->] (0,0) --
            ($(\nearrepx, \nearrepy)$)
            node[anchor=east] {$-\nabla P_i$};

        \pgfmathsetmacro{\farpa}{25};
        \pgfmathsetmacro{\farph}{.25}; 
        \pgfmathsetmacro{\farpr}{f(\farpa,\farph,\pi,\Ri,\Rdi,\ki,\xoi)}; 
        \pgfmathsetmacro{\farpx}{\farpr*cos(\farpa)};
        \pgfmathsetmacro{\farpy}{\farpr*sin(\farpa)};
        \pgfmathsetmacro{\farpd}{sqrt((\farpx)^2+(\farpy)^2)};
        \pgfmathsetmacro{\farrepx}{-.25/\farpd*\farpx};
        \pgfmathsetmacro{\farrepy}{-.25/\farpd*\farpy};

        \coordinate (farp) at (\farpx, \farpy);
        \node[circle,fill=Firebrick2,inner sep=0pt,
            minimum size=6pt, color=Firebrick2,
            label=east:{\color{Firebrick2}$x_j$}]
            at (farp) {};

        \draw[thick,color=Firebrick2,
            line cap=round, ->] (0,0) --
            ($(\farrepx, \farrepy)$)
            node[anchor=east] {$-\nabla P_j$};

        \pgfmathsetmacro{\wallhs}{hs(-90,\d,\pi,\Ri,\Rdi,\ki,\xoi)};
        \pgfmathsetmacro{\wallrepx}{0};
        \pgfmathsetmacro{\wallrepy}{\wallhs/\d};
        \draw[thick,color=gray,->] (0,0) --
            (0, \wallrepy)
            node[anchor=north east] {$-\nabla P_b$};

        \draw[very thick, color=Firebrick3,
            dashed, line cap=round, ->] (0,0)
            -- (1,0)
            node[pos=.2] (angleStart) {}
            node[anchor=north east] {$g(N_T)$};

        \pgfmathsetmacro{\repulsex}{\wallrepx+\farrepx+\nearrepx};
        \pgfmathsetmacro{\repulsey}{\wallrepy+\farrepy+\nearrepy};
        \pgfmathsetmacro{\repulsenorm}{norm(\repulsex,\repulsey)};
        \pgfmathsetmacro{\repulsescale}{gscale(\repulsex,\repulsey,3)};
        \pgfmathsetmacro{\normrepx}{\repulsex/\repulsescale};
        \pgfmathsetmacro{\normrepy}{\repulsey/\repulsescale};

        \draw[->,very thick, dashed, Firebrick3,
            line cap=round] (0,0)
            -- (\normrepx,\normrepy)
            node[anchor=south ]
            {\footnotesize $g(-\sum_k \nabla P_k)$\ \ \ \ \ \ \ };

        \pgfmathsetmacro{\resultx}{\normrepx+1};
        \pgfmathsetmacro{\resulty}{\normrepy};
        \pgfmathsetmacro{\resultscale}{gscale(\resultx,\resulty,3)};
        \draw[->,ultra thick, Firebrick4]
            (0,0) --
            (\resultx/\resultscale,
            \resulty/\resultscale )
            node[pos=.4] (angleEnd) {}
            node[anchor=south] {\footnotesize $g(g(N_T)+g(-\sum_k \nabla P_k))$};

            \node (orig) at (0,0) {};

            \draw[solid,thick,
                out=90,
                in=315,
            Firebrick4] (.25,0)
                -- ($(.4*\resultx, .4*\resulty)$)
                node[pos=.6, anchor=west] {$\theta$};
    \end{axis}

\end{tikzpicture}
    \caption{
    A pedestrian
    at the origin
    experiences repulsive
    forces $\nabla P_k$
    from the wall and two
    pedestrians $x_i,x_j$.
    As a result, 
    the GNM~\cite{pedestrians}
    shifts her orientation
    $g(N_T)$ by $\theta$
    degrees. }
    \label{fig:repulsion}
\end{figure}
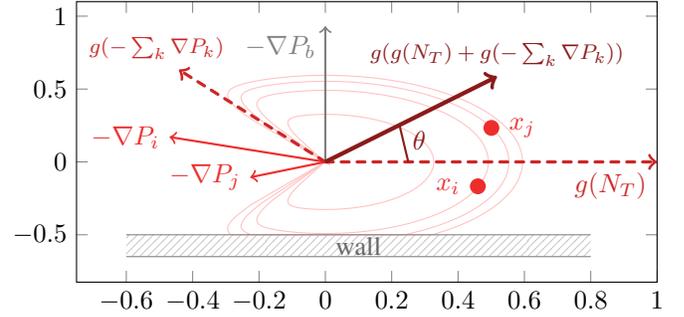

In this section we compute
the average change of orientation
$\mathbb{E}[\theta((i+k)\omega^{-1})
-\theta(i\omega^{-1})]$
as follows
\begin{enumerate}[label=\roman*)]
    \item we consider that
        pedestrians~\cite{stochastic-geometry}
        coordinates follow
        an homogeneous
        PPP with average
        rate
        $\mu$\,\textrm{ped/m\textsuperscript{2}};
    \item we compute the average force
        that pedestrians induce in
        a band of influence
        $\nabla P(H_\epsilon,
        k)$ when the pedestrian is
        $d$\,\textrm{meters} away
        from a wall
        -- see Fig.~\ref{fig:isolines}
        and Fig.~\ref{fig:repulsion}; and
    \item we average the
        influence of each
        band $H_\epsilon$
        to obtain the average change of
        orientation. 
\end{enumerate}

Following the lead
of~\cite{stochastic-vehicular},
we consider that pedestrians
coordinates are goberned
by an homogeneous PPP
$\Phi=\{x_j\}_{j\in\mathbb{N}}$
of rate $\mu$. Hence, we
know~\cite{stochastic-geometry}
the probability of having
$k$ pedestrians within the
band $H_\epsilon$ is
precisely
$\mathbb{P}(N(H_\epsilon)=k)
=(\mu|H_\epsilon|)^k/k!\cdot e^{-
\mu|H_\epsilon|}$,
with $|H_\epsilon|$ being the area
for the band of width
$\epsilon$ starting
at the isoline
$h(r)s(\theta)=H$
(see~Fig.~\ref{fig:isolines}).

However, sometimes the
pedestrian is near an
obstacle as a wall and the
area of the band
$H_\epsilon$ is clipped by the
the wall
-- see~Fig.~\ref{fig:repulsion}.
Hence the area of the band
of influence given that a wall is
$d$\,\textrm{meters} away is
\begin{equation}
    |H_\epsilon(d)|
    = \int_0^{2\pi}
    \int_{r(\theta,H)}^{r(\theta,H+\epsilon)} \rho
    \cdot
    \mathds{1}_{[-d,\infty)}(\rho\sin(\theta))\ d\rho\ d\theta
    \label{eq:area}
\end{equation}
with $r(\theta,H)\in\mathbb{R}^{+}$
a function that maps the
distance at which is the
isoline $H$ with polar
coordinates $\theta$,
$\mathds{1}_A(x)$ the
indicator function taking
value one if $x\in A$ and
zero otherwise.
We refer the reader to
\eqref{eq:radius} in
Appendix~\ref{app:radius}
for a detailed
definition of
$r(\theta,H)$.

Note that~\eqref{eq:area} only considers
the case of a wall bellow the pedestrian,
as illustrated in
Fig.~\ref{fig:repulsion}. In case there
are two walls surrounding a sidewalk
(or the start of a road and a wall), we
apply~\eqref{eq:area} in one half of the
sidewalk, and the same expression in the
other half.

Given the area of the band
$H_\epsilon(d)$, we can
obtain a closed-form
expression for the average
influence of neighbors
within such band.

\begin{lemma}[Average
    band influence]
    Given a pedestrian
    $d$ meters away
    from a wall/obstacle
    and a bounding box
    $B\subset\mathbb{R}^2$,
    the average influence
    of the band
    $H_\epsilon\subset B$ is
    \begin{multline}
        \mathbb{E}[\nabla
        P(H_\epsilon,d)]=
        \sum_{n=0}^\infty
        \sum_{k=0}^n
        \bigg(
        \mathbb{E}[
        \nabla P(H_\epsilon,
        k,d)]\\
        \cdot
        \frac{\mu^{2n}
        |H_\epsilon|^k
        |\overline{H}_\epsilon
        |^{n-k} |B|^n}{n!k!
        (n-k)!}
        e^{-2\mu|B|}
        \bigg)
        \label{eq:avg-influence}
    \end{multline}
    with $\overline{H}_\epsilon=B\setminus H_\epsilon$,
    and $\nabla P(H_\epsilon,
    k,d)$ the
    average influence
    of band $H_\epsilon$
    with an obstacle/wall
    $d$ meters away and
    with $k$ pedestrians
    inside
    \begin{multline}
        \mathbb{E}[
            \nabla P(H_\epsilon,
        k,d)]\\=
        \int_0^{2\pi}
        \frac{k\ 
        h(r(\theta,H\frac{\epsilon}{2}))
        s(\theta)
        g_\epsilon(\theta,H,d)}{2\pi}
        (\cos\theta,\sin\theta)
        \ d\theta
        \label{eq:k-avg}
    \end{multline}
    with $g_\epsilon(\theta,H,d)=
    \mathds{1}_{[-d,\infty)}
    (r(\theta,\tfrac{H+\epsilon}{2})\sin \theta)$
    \label{lemma:band}
\end{lemma}
\begin{proof}
    First we compute
    $\mathbb{E}[
    \nabla P(H_\epsilon,
        k,d)]$. Without
    loss of generality,
    we consider that the
    a pedestrian within
    the band of influence
    $H_\epsilon$ is at the
    isoline
    $H_\epsilon+\tfrac{%
    \epsilon}{2}$ -- i.e.
    at the middle -- for
    $\epsilon>0$ small
    enough. Such
    pedestrian $j$ would have
    polar coordinates
    $x_j=(r(\theta_j,H_\epsilon
    +\tfrac{\epsilon}{2}),
\theta_j)=(r_j,\theta_j)$ and its
    ocurrence would have
    probability
    $\mathbb{P}(
    \Phi=x_j|\ N(H_\epsilon)
    =k)=\tfrac{k}{2\pi}$

    According
    to~\eqref{eq:repulsion},
    the pedestrian $j$
    would induce a repulsion
    $\nabla P(\theta_j)=
    h(r_j,\theta_j)
    s(\theta_j)x_j/r_j$
    as long as $x_j$
    is not at the other
    side of the wall at
    distance $d$, i.e.
    as long as
    $g_\epsilon(\theta_j,
    H,d)=1$.
    Consequently, the average
    influence of band
    $H_\epsilon$ is
    \begin{multline}
        \mathbb{E}
        [\nabla P(H_\epsilon,
        k,d)]=
        \mathbb{E}_{\Phi}
        [\nabla P(H_\epsilon,
        k,d)|\
        N(H_\epsilon)=k]\\
        =\int_0^{2\pi}
        \nabla P(\theta_j)\ 
        \mathbb{P}(\Phi=x_j|\
        N(H_\epsilon)=k)
        \ d\theta_j
    \end{multline}
    wich results
    into~\eqref{eq:k-avg}
    in cartesian coordinates.

    Now we compute what is
    the average band
    influence resorting
    to the PPP density
    function:
    \begin{multline}
        \mathbb{E}[
        \nabla P(H_\epsilon,
        d)]\\
        =\sum_n^\infty
        \mathbb{E}
        [\nabla P(H_\epsilon,d)|\ N(B)=n]\ \mathbb{P}(N(B)=n)\\
        =\sum_n^\infty
        \sum_{k=0}^{n}
        \mathbb{E}
        [\nabla P(H_\epsilon,d)|\ N(H_\epsilon)=k, N(B)=n]\\
         \qquad\qquad\qquad \cdot\mathbb{P}(N(H_\epsilon)=k|\ N(B)=n)\ \mathbb{P}(N(B)=n)\\
        =\sum_n^\infty
        \sum_{k=0}^{n}
        \mathbb{E}
        [\nabla P(H_\epsilon,k,d)]\\
        \qquad\qquad\qquad \cdot\mathbb{P}(N(H_\epsilon)=k, N(\overline{H}_\epsilon)=n-k)\ \mathbb{P}(N(B)=n)
    \end{multline}
    which
    yields~\eqref{eq:avg-influence}
    using the PPP
    density function.
\end{proof}

Thanks to
Lemma~\ref{lemma:band}
we can 
approximate, on average,
the change of orientation.
Specifically, we compute the change
of orientation with a wall/obstacle
$d$ \textrm{meters} away.

\begin{lemma}[Average Change of Orientation]
    A pedestrian $d$\,\textrm{meters} away
    from the wall experiences an
    average change of orientation
    \begin{multline}
        \mathbb{E}\left[
            \theta((i+k)\omega^{-1})
            -\theta(i\omega^{-1})
            |\ d
        \right]\\
        =\arccos\left(\frac{
            g_1(-\nabla P_b
            -\mathbb{E}[\nabla P(H_\epsilon,
            d)])
        }{
            \|g(-\nabla P_b
            -\mathbb{E}[\nabla P(H_\epsilon,
            d)])\|
        }\right)
        -\theta(i\omega^{-1})
        \label{eq:avg-change-orient}
    \end{multline}
    with $k>0,i\geq0$;
    and $g_1(\cdot)$ the first
    coordinate of $g(\cdot)$;
    $\nabla P_b
    =(0,h(d;p_b,R_b))/d$
    the wall
    repulsion~\cite[(6)]{pedestrians}
    with $p_b,R_b$ the maximum
    and support of $h(\cdot)$.
    \label{lemma:avg-orient-change}
\end{lemma}
\begin{proof}
    For $\theta(i\omega^{-1})$ refers to the
    current orientation of the pedestrian,
    all we have to compute is the average
    orientation at time $(i+k)\omega^{-1}$
    \begin{multline}
        \mathbb{E}[\theta((i+k)\omega^{-1})|\ d]\\
        = \mathbb{E}\left[\arccos\left(
        \frac{x_1((i+k)\omega^{-1})
        -x_1((i+k-1)\omega^{-1})}{
        \|x((i+k)\omega^{-1})
        -x((i+k-1)\omega^{-1})\|
        }
        \right)\right]
        \label{eq:avg-orient}
    \end{multline}
    
    Now we obtain the average change of
    position as follows
    \begin{multline}
        \mathbb{E}[x((i+k)\omega^{-1})
        -x((i+k-1)\omega^{-1})|\ d]\\
        =x(i\omega^{-1})
        +k\omega^{-1}
        \mathbb{E}[w((i+k)\omega^{-1})]
        \mathbb{E}[N_P|\ d]\\
        -x(i\omega^{-1})
        -(k-1)\omega^{-1}
        \mathbb{E}[w((i+k-1)\omega^{-1})]
        \mathbb{E}[N_P|\ d]\\
        =\frac{\overline{w}}{
        \omega}g(-\nabla P_b -
            \mathbb{E}[\nabla P(H_\epsilon,d)])
        \label{eq:avg-change-pos}
    \end{multline}
    with the first equality holding 
    by taking a $k,k-1$ sized step in
    the Euler method and the ODE system
    in~\eqref{eq:x-der};
    $\overline{w}$ the
    average speed of a pedestrian
    -- approximately
    1.34\,\textrm{m/sec}
    \cite{pedestrians};
    and $g(\cdot)$ the
    vector normalization
    in~\cite[Appendix~1]{pedestrians}.
    Note that plugging
    \eqref{eq:avg-change-pos}
    into \eqref{eq:avg-orient}
    results into \eqref{eq:avg-change-orient}.
\end{proof}

Thanks to Lemma~\ref{lemma:avg-orient-change},
in the next section
 \S\ref{subsec:avg-rate}
we obtain the average rate of VAMs generated
due to change of orientations.

\subsection{Average Rate
of VAMs Upon Change of Orientation}
\label{subsec:avg-rate}
With the average
orientation we
now compute what is the
average rate of VAMs that
any pedestrian would generate due to changes
of orientation
-- i.e. when
$|\mathbb{E}[\theta((i+k)\omega^{-1})|\ d]-\theta(i\omega^{-1})| > \delta_\theta$
occurs.
In particular, we
approximate
$\lambda_\theta$
assuming that speed change VAMs
are negligible $\lambda_\sigma(\omega)\to0$.
Such assumption holds
--- as evidenced
in \S\ref{subsec:validation} -- due
to the tendency of having less
speed changes as the ODE warmup
increases. Notice how
$\lambda_\sigma(\omega)\to0$ as
$i_0\to\infty$
in Lemma~\ref{lemma:speed}.

Before obtaining
$\lambda_\theta$ it is worth
mentioning that the change
of orientation is computed
against the orientation
reported in the last VAM.
Consequently, if a change
of position VAM occurs before
a change of orientation
--- let us say at time
$t=3\omega^{-1}$ ---
the next check for the
orientation check is computed
as
$|\mathbb{E}[\theta(4\omega^{-1})|\ d]-\theta(3\omega^{-1})|$
which we can approximate as
$|\mathbb{E}[\theta((i+1)\omega^{-1})|\ d]-\theta(i\omega^{-1})|$.
Therefore, a VAM due to a change of orientation
must happen before a VAM due to a change
of position, which happens each
each $\omega\lceil\omega\Delta/\sigma\rceil$\,\textrm{sec}.

Consequently, we approximate the probability
of sending an orientation
change VAM at the
$k$\textsuperscript{th}
sampling period as
\begin{multline}
    p_\theta(k)
    =
    \int_{d_m}^{d_M}
    \frac{\mathds{1}_{>\delta_\theta}\left(
        |\mathbb{E}[\theta((i+k)\omega^{-1})| x]-\theta(i\omega^{-1})|
    \right)}{d_M-d_m} dx\\
    =
    \int_{d_m}^{d_M}
    \frac{\mathds{1}_{>\delta_\theta}\left(
        |\mathbb{E}[\theta(k)|\ x]-\theta(0)|
    \right)}{d_M-d_m}\ dx,
    \quad k\leq
    \left\lceil\frac{\Delta\omega}{\sigma}
    \right\rceil
\end{multline}
as long as $k$ is below the number of
samples it takes to trigger a VAM due
to a change of position -- i.e.
as long as
$k\leq\lceil\Delta\omega/\sigma\rceil$.

Knowing that 
VAMs due to position changes
restart the check of
the orientation change,
we obtain
the average rate of VAMs
due to orientation changes in the next
Lemma. 
\begin{lemma}[Average rate of orientation VAMs]
    A pedestrian with
    average speed $\sigma$
    and positioning sampling rate
    $\omega$ sends
    orientation change VAMs
    at an average rate
    $\lambda_\theta$ whose
    inverse satisfies
    \begin{equation}
        \mathbb{E}[\lambda_\theta^{-1}]=
        \frac{\omega^{-1}}{%
        1-p_{err}}
        \left(
            p_1+
            \frac{I_\Delta p_0 p_{err}}{1-p_{err}}
        \right)
    \end{equation}
    with $I_\Delta=
    \lceil\Delta\omega/\sigma\rceil$ and
    $p_0,p_1,p_{err}$ 
    as defined in
    \mbox{\eqref{eq:p0}-\eqref{eq:perr}}.
    \label{lemma:orient-rate}
\end{lemma}
\begin{proof}
    We first compute what
    is the probability
    of a orientation VAM
    occurring at the
    $k$\textsuperscript{th}
    sampling period
    \begin{equation}
        \mathbb{P}(
        \lambda_\theta^{-1}=
        k\omega^{-1})
        = 
        p_\theta(k)
            \prod_{m=1}^{k-1}
            (1-p_\theta(m)),
        \quad k<I_\Delta
    \end{equation}
    with $I_\Delta=\lceil           \Delta\omega/\sigma
        \rceil$
    the number of sampling
    periods $\omega^{-1}$
    required to trigger a
    position VAM.
    
    Next, we compute the
    probability of an
    orientation VAM
    occuring
    at the
    $k$\textsuperscript{th}
    sampling period
    after the
    $j$\textsuperscript{th}
    position VAM
    \begin{multline}
        \mathbb{P}\left(
        \lambda_\theta^{-1}=
        \left(
            jI_\Delta
            +k
        \right)
        \omega^{-1}
        \right)\\
        = 
        \left(
            \prod_{m=1}^{%
            I_\Delta}
            (1-p_\theta(m))
        \right)^j
        \mathbb{P}(
        \lambda_\theta^{-1}
        =k\omega^{-1})
        ,
        \quad k<I_\Delta
    \end{multline}

    Then we compute the
    average rate as
    \begin{equation}
        \mathbb{E}[\lambda_\theta^{-1}]
        =
        \sum_{j=0}^\infty
        \sum_{k=1}^{I_\Delta-1}
        \frac{%
        jI_\Delta+k}{\omega}
        \mathbb{P}\left(
        \lambda_\theta^{-1}=
        \left(
            jI_\Delta
            +k
        \right)
        \omega^{-1}
        \right)
        \label{eq:closed-def}
    \end{equation}
    
    If we call
    \begin{align}
        p_0&=\sum_{k=1}^{I_\Delta-1}
        \mathbb{P}
        (\lambda_\theta^{-1}
        =k\omega^{-1})\label{eq:p0}\\
        p_1&=\sum_{k=1}^{I_\Delta-1}
        k\mathbb{P}
        (\lambda_\theta^{-1}
        =k\omega^{-1})\\
        p_{err}&=
        \prod_{m=1}^{I_\Delta}
        (1-p_\theta(m))\label{eq:perr}
    \end{align}
    expression~\eqref{eq:closed-def}
    becomes
    \begin{multline}
        \mathbb{E}[{\lambda_\theta^{-1}}]=
        \omega^{-1}
        \sum_{j=0}^\infty
        \sum_{k=1}^{I_\Delta-1}
        (jI_\Delta+k)
        p_{err}^j
        \mathbb{P}(\lambda_\theta^{-1}
        =k\omega^{-1})\\
        = \omega^{-1}
        \sum_{j=0}^\infty
        p_{err}^j
        \left(
        jI_\Delta
        p_0 + p_1
        \right)\\
        =\omega^{-1}
        \left(
        \frac{I_\Delta p_0 p_{err}}{(1-p_{err})^2}
        +\frac{p_1}{1-p_{err}}
        \right)
    \end{multline}
\end{proof}

\section{Pedestrian IPG minimization}
\label{sec:minimize}

In this section we describe
which is the adequate
sampling rate $\omega$
to minimize the average
Inter Packet Gap
(IPG) of VAMs.
Decreasing the IPG results
into increasing the rate at
which VAMs are received,
hence resulting into fresh
information regarding the
pedestrians' positions.
In particular, we are
interested into decreasing
the IPG of VAMs belonging
to Profile~2 VRUs~\cite{etsi-va}, i.e.
the IPG between pedestrian
VAMs. From now on we will
refer to the pedestrian
VAMs' IPG as pIPG.

To obtain the average pIPG
we resort to the rate
at which pedestrians send
traffic. Each pedestrian
sends an average rate of
$\lambda(\omega)=
\lambda_\Delta
+\lambda_\sigma(\omega)
+\lambda_\theta(\omega)$\,\textrm{VAM/sec}
that we obtain
using~\eqref{eq:lambda},
Lemma~\ref{lemma:speed},
and Lemma~\ref{lemma:orient-rate};
respectively.
On top, other
VRUs (e.g. bikes)
and vehicles (e.g. cars)
send their corresponding
VAM and CAM messages over
the wireless channel.
We refer to $\lambda_b,
\lambda_c$ as the average
rate of VAMs of other
VRUs as bikes, and the
average rate of CAMs of
vehicles as cars -- thus,
the subscripts.
Similarly, we refer to
$n_p$ as the number of
pedestrians sending VAMs,
$n_b$ as the number of
bikes (other VRUs) sending
VAMs, and $n_c$ as the 
number of vehicles sending
CAMs.

Consequently, the wireless
channel foresees the
following average
aggregated rate of
VAMs and CAMs
\begin{equation}
    \Lambda(\omega)=
    \lambda_b n_b
    +\lambda_v n_v
    +\lambda(\omega) n_p
    = \Lambda_v+\Lambda_b + n_p
    \lambda(\omega)
    \label{eq:global-rate}
\end{equation}
with $\Lambda_v,\Lambda_b$
the total rate of other
VRUs and vehicles.
Note that it is possible
to estimate both
$n_b,n_v$ and
$\lambda_b,\lambda_v$ either
from an RSU, or with
estimations carried over
time as the device
listens to the wireless
channel. Even if a VAM/CAM
is lost due to collissions,
it is possible to infer
the rate checking the
identifiers of the
successfully received
VAMs/CAMs.

To derive the pIPG we
resort to the following
lemma.

\begin{lemma}[VRU average
    pIPG]
    And VRU with checking
    frequency $\omega$,
    speed $\sigma$,
    checking distance
    $\Delta$ experiencing
    a PDR $p(\Lambda(\omega))$ foresees an
    average pIPG:
    \begin{equation}
        \mathbb{E}[\text{\normalfont pIPG}]
        =\frac{1}{n_p\lambda(\omega)p(\Lambda(\omega))}
    \label{eq:avg-pipg}
    \end{equation}
    with $n_p$ the number of
    active pedestrian
    VRUs,
    $\lambda(\omega)$ their
    average VAM rate,
    and $p(\Lambda(\omega))$
    the PDR in the wireless
    medium due to the
    global VAM/CAM rate
    upon a sampling
    frequency $\omega$.
    \label{lemma:pipg}
\end{lemma}
\begin{proof}
    The wireless
    channel foresees
    an aggregated rate of
    $n_p\lambda(\omega)$\,\textrm{VAM/sec}
    from pedestrians.
    Hence, the average time
    ellapsing between
    pedestrian VAMs is
    $\tfrac{1}{n_p\lambda(\omega)}$\,\textrm{sec}.
    However,
    it may happen that
    a pedestrian VAM is
    lost with probability
    $1-p(\Lambda(\omega))$,
    with the latter term
    referring to the
    PDR due to the
    aggregate VAM rate in
    the channel, considering
    that pedestrians have
    a sampling rate
    $\omega$.
    Consequently, the
    probability of
    $i-1$ consecutive
    pedestrian VAMs
    being lost in the
    wireless channel,
    and having an
    i\textsuperscript{th}
    VAM transmission
    success has
    probability
    $p(\Lambda(\omega))
    [1-p(\Lambda(\omega))]^{i-1}$.
    Therefore, the
    average IPG is computed
    as:
    \begin{multline}
        \mathbb{E}[\text{pIPG}]
        =\sum_{i=1}^\infty
        \frac{i}{n_p
        \lambda(\omega)}
        \ p(\Lambda(\omega))[1-p(\Lambda(\omega))]^{i-1}\\
        = \frac{1}{%
        n_p\lambda(\omega)
        p(\Lambda(\omega))}
        \label{eq:proof-ipg}
    \end{multline}
    with $\tfrac{i}{n_p\lambda(\omega)}$
    the time ellapsed between
    first and
    i\textsuperscript{th}
    VAM sent to the channel.
\end{proof}

To find the optimum sampling
rate we check which
pedestrian VAM rate
$\lambda(\omega)$ minimizes
the average pIPG.

\begin{problem}[VAM pIPG minimization]
    Given $n_b$ bikes with average
    VAM rate $\lambda_b$,
    $n_c$ cars with average VAM rate
    $\lambda_c$, and $n_p$ pedestrians;
    solve
    \begin{align}
        \min_{\omega} &\quad \mathbb{E}[\textnormal{pIPG}]\\
        \text{s.t.:} & \quad \omega\leq\omega_{\max}
    \end{align}
    \label{problem:min-pipg}
    with $\omega_{\max}$ 
    the maximum positioning sampling frequency.
\end{problem}

To solve Problem~\ref{problem:min-pipg}
we take the derivative
of~\eqref{eq:avg-pipg} 
denoting
$\lambda=\lambda(\omega)$, i.e.:
\begin{equation}
    \frac{d}{d\lambda}
    \mathbb{E}[\text{pIPG}]
    =\frac{-1}{n_p(\lambda
    p(\lambda))^2}
    \left[ p(\lambda)
    +\lambda p'(\lambda)
    \right]
    \label{eq:pipg-der}
\end{equation}
with $p(\lambda)
=p(\Lambda(\omega))$
to ease notation given
the dependancy of
$\Lambda(\omega)$ on
$\lambda(\omega)$
--
see~\eqref{eq:global-rate}.

From~\eqref{eq:pipg-der}
we see that a local
optimum is found at
\mbox{$\lambda_0:
p(\lambda_0)=-\lambda_0
p'(\lambda_0)$}.
We know the PDR
$p(\lambda)\in[0,1]$
is a monotonically
decreasing function on
$\lambda$ -- i.e. the
transmission success
probability decreases
as the rate $\lambda$
increases -- and
$p(0)=1$.
But $p(\lambda)$ depends on the
used radio technology -- e.g. NR or 802.11p --,
and it may be that $\lambda_0$ has multiplicity
greater than one, i.e. that there are
multiple local minimums.

We take
the second derivative of
the average pIPG to gain
further understanding
\begin{equation}
    \frac{d^2}{d\lambda^2}
    \mathbb{E}[\text{pIPG}]
    =\frac{2\left[
        p(\lambda)
        +\lambda p'(\lambda)
    \right]^2}{n\left(
    \lambda p(\lambda)
    \right)^3}
    -\frac{2p'(\lambda)
    +n\lambda p''(\lambda)}{n(\lambda p(\lambda))^2}
    \label{eq:pipg-2der}
\end{equation}

For we do not know how
the second derivate
of the PDR is, we cannot tell
whether
$p''(\lambda)>0, \forall
\lambda$
to fully determine the convexity
of $\tfrac{d^2}{d\lambda^2}
\mathbb{E}[\text{pIPG}]$, hence,
to conclude that
$\lambda_0$ has multiplicity one
and there exists an unique
global minimum.

To gain further understanding on
the second derivative of the
expected pIPG, in the next section
we study $p(\lambda)$ for 802.11p,
and solve Problem~\ref{problem:min-pipg}.
Towards the end of the next section
we also explain how to adapt the proposed
802.11p solution to other technologies
as 802.11bd and C-V2X.

\subsection{802.11p
Pedestrian IPG
Minimization}
\label{subsec:80211p}

In this section we resort to the
PDR  expression $p(\Lambda(\omega))$
of the 802.11p node model
in~\cite[{\S}III-A]{BaiocchiAoI},
and find the optimal $\omega$ to
solve Problem~\ref{problem:min-pipg}.
The node model in~\cite{BaiocchiAoI}
takes two assumptions:
\begin{enumerate}[label=A\arabic*)]
    \item 802.11p devices
    generate packets
    (VAMs/CAMs) at a Poissonian\label{assumption:1}
    rate; and
    \item 802.11p devices
        have an homogeneous
        VAM/CAM rate
        that we obtain as
        an equal split
        of the aggregate
        rate
        in~\eqref{eq:global-rate}
        -- i.e. each device
        has an average
        rate
        $\Lambda(\omega)/n$.\label{assumption:2}
\end{enumerate}
Both assumptions are taken
for the sake of analytical
tractability.
Although
assumption
\ref{assumption:1}
does not
hold -- see~\cite[Figure~B.4]{histogram} --,
it serves us to take
a conservative approximation
on the optimum $\omega$
due to the long tail
of the exponential
inter-arrival times of
Poissonian processes.
Assumption
\ref{assumption:2}
is neither
realistic for
cars generate CAMs
at a higher rate than
pedestrian VAMs.
However, assumptions
\ref{assumption:1} and
\ref{assumption:2}
hold upon sufficiently
large VAMs and CAMs in
802.11p, for the aggregation
of arrival processes results
into a Poissonian arrival
process according to 
the Palm-Khintchine
theorem~\cite{p-k}.
Nevertheless, later in
 \S\ref{subsec:simulation}
we investigate scenarios
with small/large number
of VAM/CAMs to highlight
the drawback/advantages
of assumptions
\ref{assumption:1} and
\ref{assumption:2}.

The node model obtains a
recurrence
expression~\cite[(15)]{BaiocchiAoI}
for the probability $\tau$ of
an 802.11p node transmitting
at a virtual time slot
that consists of
\textit{an idle back-off
slot, or an idle back-off
slot followed by a 
transmission}
-- see~\cite[{ \S}II-C]{BaiocchiAoI}.
Specifically, the PDR
is obtained as
\begin{equation}
    p(\Lambda(\omega))
    = (1-\tau)^{n-1}
    \label{eq:pdr-node}
\end{equation}
with $\tau$ depending
on the aggrefated
rate of VAMs and CAMs
$\Lambda(\omega)$.

While using the expressions
of~\cite{BaiocchiAoI} we
found a typo in the
manuscript that we correct
to obtain the PDR term
$\tau$
-- see~Appendix~\ref{app:correction} for further
details.

For $\tau$ does not have
either a closed-form
expression
-- it is found through
fixed-point
iteration~\cite[{\S}III-B]{BaiocchiAoI} --
we cannot either conclude
whether the local optimum
for the average pIPG
near $\lambda=0$ is a
global optimum. However,
we see through numerical
inspection that the
PDR~\eqref{eq:pdr-node}
has two concave regions,
hence, two candidate
global minimums.

\begin{figure}[t]
    \input{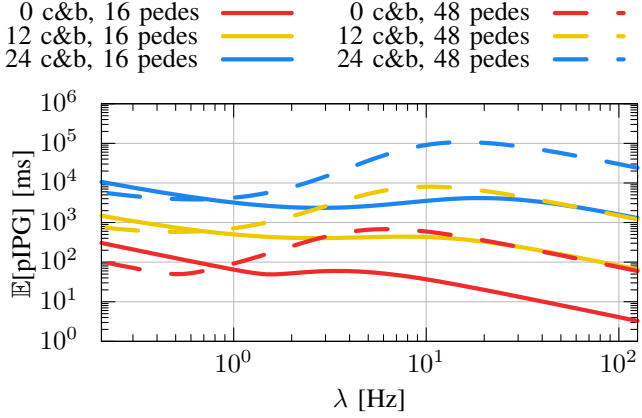}
    \vspace{-1em}
    \caption{Average
    pIPG vs.
    the pedestrian VAM
    rate for densities
    $\mu=0.01$\,\textrm{ped/m\textsuperscript{2}}
    (16~ped),
    $\mu=0.03$\,\textrm{ped/m\textsuperscript{2}}
    (32~ped), and
    equally increasing
    amount
    of cars and bikes
    (c\&b)
    $n_b=n_v$.
    CAM/VAM rates are
    $\lambda_c=3$
    and
    $\lambda_b=1$\,\textrm{Hz}.
    }
    \label{fig:valleys}
\end{figure}

In Fig.~\ref{fig:valleys}
we depict how the pedestrian
VAM rate $\lambda$
influences the average
pIPG as the pedestrians,
bikes and cars increase.
Results are obtained plugging
the PDR expression from the
node model~\eqref{eq:pdr-node}
into the average pIPG expression
in~\eqref{eq:avg-pipg}.

Fig.~\ref{fig:valleys}
evidences the existence of two
concave regions.
Consequently, gradient descends from
$\lambda=0$ leads to a global optimum
only in certain scenarios
-- e.g. for the 24 bikes \&
cars and 16 pedestrians
in Fig.~\ref{fig:valleys}.
In other scenarios,
gradient descends starting
from $\lambda=0$ get stuck in
the left local optimum, and do
not reach the global minimum
at the right
-- e.g. for the scenario
without bikes and cars with
16 pedestrians
in Fig.~\ref{fig:valleys}.

Motivated by the numerical
evidences of the average pIPG
-- see Fig.~\ref{fig:valleys} --,
we conjecture that it
is possible to solve
Problem~\ref{problem:min-pipg}.
Specifically, we propose 
Algorithm~\ref{alg:search} to obtain
the sampling rate $\omega$ 
that minimizes the average pedestrian
IPG.

\algrenewcommand\algorithmicrequire{\textbf{Input:}}
\algrenewcommand\algorithmicensure{\textbf{Output:}}
\begin{algorithm}[t]
\caption{802.11p pedestrian
sampling rate
search}\label{alg:search}
\begin{algorithmic}[1]

\Require $n, n_p,
\Lambda_b,\Lambda_v,
\left\{\lambda_\theta(
\omega_i)\right\}_{i=0}^N$
\Ensure $\omega^*$


\State $\lambda=\lambda_{\max}$

\State $\text{pIPG}_0=-10$

\Comment Invoke
Algorithm~\ref{alg:pipg}
as pIPG()
\While{pIPG($n,n_p,\Lambda_b,
 \Lambda_v, \lambda)$
 $>\text{pIPG}_0$}\label{line:max}

 \State $\text{pIPG}_0$
 =pIPG($n,n_p,\Lambda_b,
 \Lambda_v, \lambda)$

 \State $\lambda=\lambda+
 \varepsilon_\lambda$

\EndWhile

\State $\lambda_l$=
Brent$\left(
    (\lambda_{\min}
        ,\lambda),
\text{pIPG} \right)$

\State $\lambda_r$=
Brent$\left(
    (\lambda,
        \lambda_{\max}
        ),
\text{pIPG} \right)$

\State $\lambda_{\min}
=\argmin_{\lambda\in
\{\lambda_l,\lambda_r\}}
\left\{
    \text{pIPG}(n,n_p,
    \Lambda_b,\Lambda_v,
    \lambda)
\right\}$\label{line:global}

\State
$\omega^*=\argmin_{\omega}
\left\{\left|
\lambda_{\min}-\lambda_\Delta
-\lambda_\sigma-\lambda_\theta(\omega)
\right|\right\}
$

\end{algorithmic}
\end{algorithm}

\begin{algorithm}[t]
\caption{Average 802.11p pIPG}\label{alg:pipg}
\begin{algorithmic}[1]

\Require $n, n_p,
\Lambda_b,\Lambda_v$
\Ensure pIPG


\State $\Lambda=\Lambda_v
+\Lambda_b+n_p\lambda$

\Comment Find PDR $\tau$
using \cite[(15)]{BaiocchiAoI} fixed point iteration
\State $\tau=$
fixed\_point\_iteration($\tau_0=0$,
rate=$\Lambda/n$,
devices=$n$)\label{line:iter}

\State $p(\Lambda)
=(1-\tau)^{n-1}$\label{line:pdr}

\State $\text{pIPG}
=[n_p\lambda
p(\Lambda)]^{-1}$

\end{algorithmic}
\end{algorithm}

The basic idea of
Algorithm~\ref{alg:search}
is to look for the
pIPG local maximum
starting near large values of
$\lambda$.
Then, it triggers two
bounded gradient descends
for small and
large values of $\lambda$
to obtain the two local
minimums shown in~Fig.~\ref{fig:valleys}.

First,
we look for the $\lambda$
corresponding to the
local maximum of the
average pIPG
-- see~line~\ref{line:max}.
The pIPG is obtained using
Algorithm~\ref{alg:pipg},
which uses
\cite{BaiocchiAoI}~node model
fixed point iteration
to find the PDR
-- Algorithm~\ref{alg:pipg}~line~\ref{line:iter} --
and return the corresponding
average pIPG.

Second, we issue two
bounded minimizations
using the Brent
method~\cite{otherbrent,brent2013algorithms}
to the left and right
of the local maximum
$\lambda$. That is,
we issue bounded minimization
within the intervals
$(\lambda_{\min},\lambda)$
and $(\lambda,
\lambda_{\max})$
to obtain the left and right local
minimums $\lambda_l,
\lambda_r$; respectively.
The global minimum
$\lambda_{\min}$ is
obtained checking which
$\lambda_l,\lambda_r$
results into a
smaller average pIPG
-- line~\ref{line:global}.

Finally, we look for
the pedestrian sampling
rate $\omega$ resulting in
a rate
$\lambda=\lambda_\Delta
+\lambda_\sigma
+\lambda_\theta(\omega)$
that is close to
$\lambda_{\min}$.
In such search we leverage
pre-computed estimations
of $\lambda_\theta(\omega_i)$
for different sampling
rates
$\omega_0,\omega_i,
\ldots,\omega_N$.
We also resort to
pre-computed
estimations of
$\lambda_\Delta,
\lambda_\sigma$.

Note that our approach is adaptable to
technologies as 802.11bd and C-V2X.
First, it is necessary to obtain a
closed-form expression or interpolation
for the PDR $p(\Lambda(\omega))$ and plug
it into Algorithm~\ref{alg:pipg}
line~\ref{line:pdr}.
Then, Algorithm~\ref{alg:search}
should be adapted to consider the
multiplicity of the solution to
$\tfrac{d}{d\lambda}
\mathbb{E}[\text{pIPG}]=0$ in the considered
technology -- e.g. 802.11bd.
If the multiplicity is one, a gradient
descend on $\lambda$ would suffice,
otherwise, it is necessary to study the
shape of $\mathbb{E}[\text{pIPG}]$ vs
$\lambda$
(see~Fig.~\ref{fig:pipg-vs-lambda}).
For example, if 
$\tfrac{d}{d\lambda}
\mathbb{E}[\text{pIPG}]=0$ has multiplicity
three, it is necessary to issue three
Brent searches (rather than the two
issued in Algorithm~\ref{alg:search}
for 802.11p)
and compare which of the three solutions
yields the minimum $\mathbb{E}[\text{pIPG}]$.

\section{Results}
\label{sec:results}

In this section we:
validate the pedestrian VAM
characterization (\S\ref{subsec:validation});
and evaluate Algorithm~\ref{alg:search}
through numerical and simulation
experiments (\S\ref{subsec:numerical}
and \ref{subsec:simulation},
respectively).

\subsection{Pedestrian VAM Rate Validation}
\label{subsec:validation}

In this section we
validate whether the
estimations specified in
\S\ref{sec:model}--\ref{sec:orient} for
$\lambda_\Delta,
\lambda_\sigma,
\lambda_\theta$
hold. For the VAMs due
to changes of position
$\lambda_\Delta$ we assume
a constant speed for its
approximation, hence, we
resort to the expression
in~\eqref{eq:lambda}.
In the case of the VAMs due
to speed changes we leverage
Lemma~\ref{lemma:speed}
using an ODE warmup of
10\,\textrm{sec} --
i.e. we take
$\lambda_\sigma\simeq
\lambda_\sigma(i_0)$
with $i_0=10$.
The VAMs due to orientation
changes are approximatted
using the average expression
from Lemma~\ref{lemma:orient-rate}.

\begin{figure*}[t]
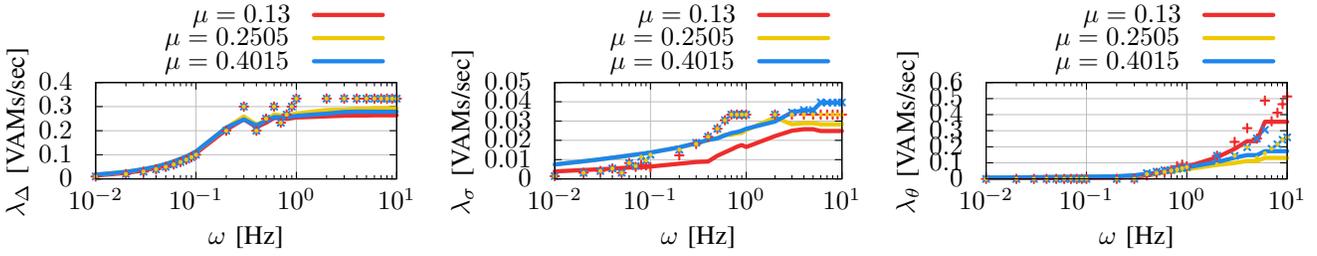

   \begin{subfigure}[b]{0.32\textwidth}
       \input{fig/lambdas_position}
   \end{subfigure}
   \begin{subfigure}[b]{0.32\textwidth}
       \input{fig/lambdas_speed}
   \end{subfigure}
   \begin{subfigure}[b]{0.32\textwidth}
       \input{fig/lambdas_orient}
   \end{subfigure}
   \vspace{-1em}
   \caption{Average VAM
   rate due to changes
   of position
   $\lambda_\Delta$,
   change of speed
   $\lambda_\sigma$ and
   change of orientation
   $\lambda_\theta$.
   Lines represent
   results obtained through
   Vadere~\cite{vadere}
   simulations and markers
   represent
   results from our
   theoretical
   approximations
    \S\ref{sec:model}--\ref{sec:orient}.}
   \label{fig:validation}
\end{figure*}

We consider that pedestrians have an average
speed of $\sigma=1.34$\,\textrm{m/sec}
and walk along the two
sidewalks of a street of
2\,\textrm{km}. We consider
three different densities
of $\mu=0.13,0.2505,0.4515$\,\textrm{ped/m\textsuperscript{2}}
and test different positioning
sampling rates in the
range
$10^{-2}\,\textrm{Hz}\leq\omega\leq\omega_{\max}$.
In particular, we choose
$\omega_{\max}=10\,\textrm{Hz}$ because
($i$) that is the maximum sampling rate
of high precission
commercial devices~\cite[12]{gpscomercial};
($ii$) we observe empirically that
pedestrian VAM rates do not increase
with $\omega\geq10$\,\textrm{Hz}; and
($iii$) ETSI~\cite[6.2]{etsi-va}
imposes a minimum positioning
sampling rate of 10\,\textrm{Hz}.
Pedestrians travel in
opposite
directions in both
sidewalks, hence leading
to situations in which
pedestrians have to avoid
people coming towards
them in the opposite
direction.

Fig.~\ref{fig:validation}
shows how
VAM rates increase with
the positioning sampling rate
$\omega$. Clearly, having
higher sampling rates
increases the chances
of detecting changes of
position, speed, and
orientation -- thus
increasing the corresponding
VAM rates
$\lambda_\Delta,
\lambda_\sigma,
\lambda_\theta$.
While the VAM rate is
monotonically increasing
on $\omega$ for speed
and orientation changes
-- see~Fig.~\ref{fig:validation}
middle and right --,
the VAM rate due to
changes of position
present notorious
oscillations when
$\omega\in[10^{-1},10^{0}]$
-- see
Fig.~\ref{fig:validation}
left.
The oscillating behaviour
is inline with the
saw-teeth curves already
presented in
Fig.~\ref{fig:saw}, and
Vadere simulations
-- points in
Fig.~\ref{fig:validation} --
are inline with the
modelled oscillating
behaviour~\eqref{eq:lambda}
for $\lambda_\Delta$.

Regardless the event
triggering the VAMs,
Fig.~\ref{fig:validation}
evidences that the VAM
rates stabilize for
large enough
sampling rates $\omega$.
Specifically,
$\lambda_\Delta$ stabilizes
for $\omega\geq1$\,\textrm{Hz};
$\lambda_\sigma$
around
for $\omega\geq3$\,\textrm{Hz}; and
$\lambda_\theta$
at sampling rates
$\omega\geq5$\,\textrm{Hz}.
Additionally,
we observe that the
stabilization of
$\lambda_\sigma$
(speed changes)
depends on the
pedestrian density
$\mu$, for it is not
as easy to e.g.
start running when
streets are empty than
when they are full.

\begin{table}[t]
    \centering
    \caption{VAM
    approximations'
( \S\ref{sec:model}--\ref{sec:orient}) errors.}
    \label{table:err}
    \begin{tabular}{c c | c c c c }
        \toprule
        & & \multicolumn{4}{c}{\textbf{Error Percentile} $|\lambda_{*}-\hat{\lambda}_{*}|$}\\
        \textbf{Density}
        & \textbf{VAM Trigger}
        & \textbf{$25$}
        & \textbf{$50$}
        & \textbf{$75$}
        & \textbf{$100$}\\ \midrule
$\mu=0.13$ & Distance $\lambda_\Delta$ & 0.006 & 0.018 & 0.070 & 0.078 \\
$\mu=0.2505$ & Distance $\lambda_\Delta$ & 0.012 & 0.026 & 0.040 & 0.062 \\
$\mu=0.4015$ & Distance $\lambda_\Delta$ & 0.012 & 0.021 & 0.054 & 0.072 \\
$\mu=0.13$ & Speed $\lambda_\sigma$ & 0.002 & 0.008 & 0.013 & 0.000 \\
$\mu=0.2505$ & Speed $\lambda_\sigma$ & 0.004 & 0.005 & 0.006 & 0.010 \\
$\mu=0.4015$ & Speed $\lambda_\sigma$ & 0.002 & 0.006 & 0.008 & 0.000 \\
$\mu=0.13$   & Orientation $\lambda_\theta$ & 0.005 & 0.006 & 0.016 & 0.157 \\
$\mu=0.2505$ & Orientation $\lambda_\theta$ & 0.008 & 0.013 & 0.020 & 0.128 \\
$\mu=0.4015$ & Orientation $\lambda_\theta$ & 0.007 & 0.012 & 0.022 & 0.132 \\
    \bottomrule
    \end{tabular}
\end{table}

Lastly, it is worth remarking
that our analytical
approximations for
$\lambda_\Delta,
\lambda_\sigma,
\lambda_\theta$ stay
close to the results obtained
through simulation
-- see~Fig.~\ref{fig:validation}
and Table~\ref{table:err}.
Indeed, our analytical
approximations are
conservative for sampling
rates $\omega>10^{-1}$\,\textrm{Hz},
i.e. simulations yield
smaller VAM rates.
In other words, our
analytical expressions for
$\lambda_\Delta,
\lambda_\sigma,
\lambda_\theta$
are conservative, yet
accurate if the positioning is
checked --- at least --- every
10\,\textrm{sec}.

\subsection{Numerical Results}
\label{subsec:numerical}

\begin{figure*}[t]
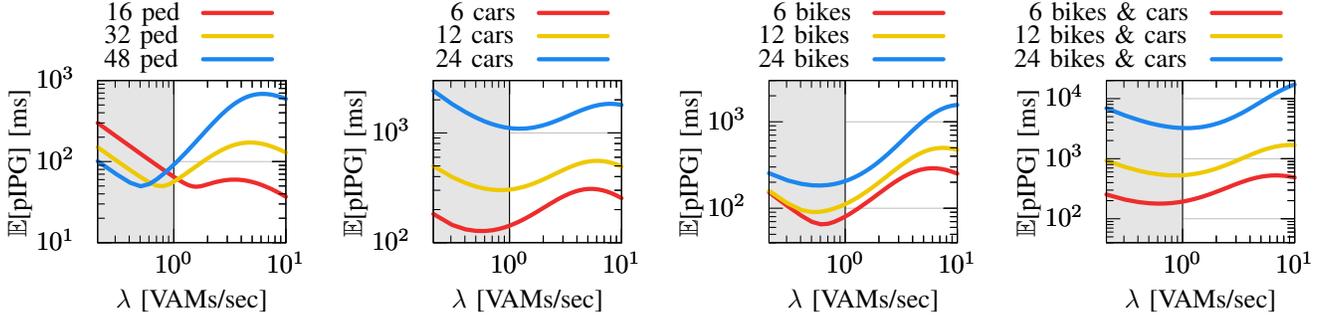

   \begin{subfigure}[b]{0.24\textwidth}
       \input{fig/only-pedestrians}
   \end{subfigure}
   \begin{subfigure}[b]{0.24\textwidth}
       \input{fig/pedestrians-cars}
   \end{subfigure}
   \begin{subfigure}[b]{0.24\textwidth}
       \input{fig/pedestrians-bikes}
   \end{subfigure}
   \begin{subfigure}[b]{0.24\textwidth}
       \input{fig/pedestrians-bikes-cars}
   \end{subfigure}
   \vspace{-1em}
   \caption{Average
    pIPG
    (y-axis)
    vs. their VAM rate
    (x-axis)
    in a sub-urban scenario.
    In the left we
    consider only pedestrians.
    The remaining scenarios
    consider 32 pedestrians
    and bikes (mid-left)
    or cars (mid-right).
    The scenario on the
    right consider 32
    pedestrians with $N$
    bikes and $N$ cars.
    Gray areas highlight
    the pedestrian
    VAM rate range.
    }
    \label{fig:pipg-vs-lambda}
\end{figure*}

\begin{figure*}[t]
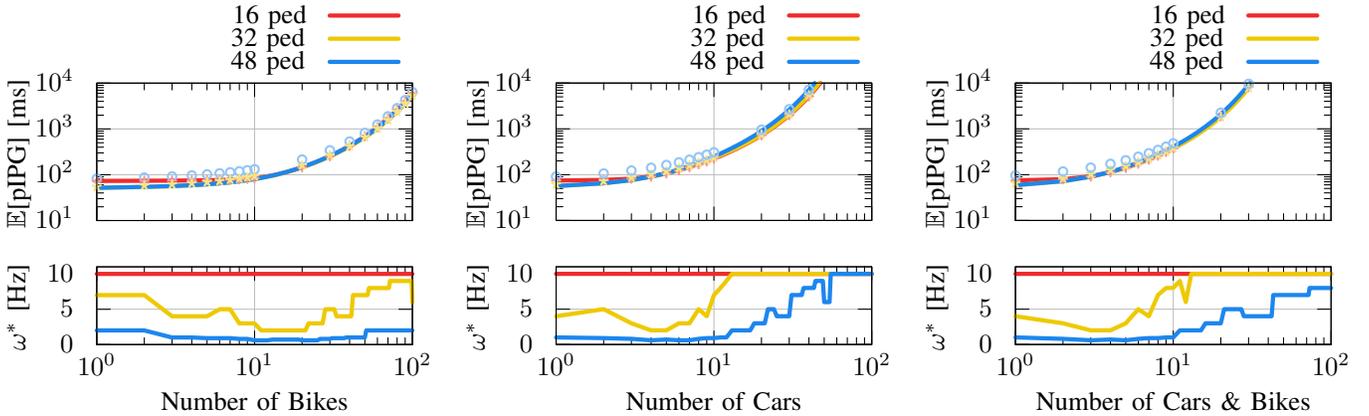

   \begin{subfigure}[b]{0.32\textwidth}
       \input{fig/bike-increase.tex}
   \end{subfigure}
   ~
   \begin{subfigure}[b]{0.32\textwidth}
       \input{fig/car-increase.tex}
   \end{subfigure}
   ~
   \begin{subfigure}[b]{0.32\textwidth}
       \input{fig/car-bike-increase.tex}
   \end{subfigure}
   \vspace{-2em}
   \caption{Optimal positioning sampling frequency
       $\omega^*$ (bottom)
       and their average pedestrians IPG 
       (top). Markers on top correspond to
       the pedestrian IPG experienced
       using ETSI~\cite{etsi-va} sampling
       frequency $\omega=10$\,\textrm{Hz}.
       Results are obtained using 
       \cite{BaiocchiAoI}~802.11p
       node model (bottom) and
       Algorithm~\ref{alg:search}.}
    \label{fig:optimal_sampling}
\end{figure*}


In this section we validate the
applicability of Algorithm~\ref{alg:search}
through numerical evaluation.
Specifically, we evaluate the performance
of Algorithm~\ref{alg:search} using
the 802.11p node model~\cite{BaiocchiAoI},
i.e. we consider that devices
obey the exponential
assumptions~\ref{assumption:1}
and \ref{assumption:2}
in \S\ref{subsec:80211p}.

In the experiments we consider an increasing
amount of bikes $n_b=6,12,24$ in
a street of $2$\,\textrm{km}
--- as in \S\ref{subsec:validation}.
Each bike as an average VAM rate of
$\lambda_b=1$\,\textrm{Hz}
\cite{etsiNewDcc}.
Similarly, we consider an increasing
amount of cars $n_c=6,12,24$ in the
$2$\,\textrm{km} street, each with
an average CAM
rate of $\lambda_c=3$\,\textrm{Hz}.
Note that the Decentralized Congestion
Control (DCC) clips the CAM rate
to at most $3$\,\textrm{Hz}
\cite{Amador2020}.

Pedestrians VAM rate $\lambda(\omega)$
will depend on
the optimal positioning sampling frequency
$\omega^*$ found by Algorithm~\ref{alg:search}
based on the pedestrian
density $\mu$ in the sidewalks
-- see~Fig.~\ref{fig:validation}.
For we consider the densities
in \S\ref{subsec:validation},
i.e. densities of
$\mu=0.13,0.2505,0.4015$\,\textrm{ped/m\textsuperscript{2}}
that correspond to
$n_p=16,32,48$ pedestrians in the
considered $2$\,\textrm{km} street.
Given the pedestrian VAM rate
characterizations from
 \S\ref{sec:model}--\ref{sec:orient},
we know that the maximum
VAM rate remains below $1$\,\textrm{Hz}
because
$\lambda_\Delta+\lambda_\sigma
+\lambda_\theta<1$\,\textrm{Hz} at
the maximum positioning sampling frequency
$\omega_{\max}=10$\,\textrm{Hz}
-- see~Fig.~\ref{fig:validation}
results.

Fig.~\ref{fig:pipg-vs-lambda}
illustrates the two local minimums that
the average pIPG presents as the
pedestrian VAM rate $\lambda$ varies
in the following scenarios:
(i) only pedestrians
Fig.~\ref{fig:pipg-vs-lambda}~(left);
(ii) 32 pedestrians with only bikes
Fig.~\ref{fig:pipg-vs-lambda}~(mid-left);
(iii) 32 pedestrians with only cars
Fig.~\ref{fig:pipg-vs-lambda}~(mid-right); and
(iv) 32 pedestrians with cars and bikes
Fig.~\ref{fig:pipg-vs-lambda}~(right).
The gray areas highlight the minimum
and maximum VAM rate generated by pedestrians
due to the maximum positioning sampling rate
limitation of $\omega_{\max}=10$\,\textrm{Hz}.

If only pedestrians are present in the street,
having more pedestrians leads to a smaller
average pIPG
--- see~Fig.~\ref{fig:pipg-vs-lambda}~(left).
Clearly, more pedestrians lead to more
VAMs, hence to smaller time ellapsed between
two pedestrian VAMs.
However, the increase of bikes or/and cars 
--- see~Fig.~\ref{fig:pipg-vs-lambda}~(mid-left)
to Fig.~\ref{fig:pipg-vs-lambda}~(right) ---
lead to a higher average pIPG.
High number of bikes and cars result
into more packet collisions in the 802.11p
channel, hence pedestrian VAMs are prone
to fail and the average pIPG increases.

Motivated by the insights of
Fig.~\ref{fig:pipg-vs-lambda},
we evaluate Algorithm~\ref{alg:search}
in scenarios with: ($i$) increasing number
of bikes
Fig.~\ref{fig:optimal_sampling}~(left);
($ii$) cars
Fig.~\ref{fig:optimal_sampling}~(middle);
and ($iii$) increasing number
of bikes and cars
Fig.~\ref{fig:optimal_sampling}~(right).
Namely, we compare the average pIPG
in the 802.11p channel using ETSI
positioning sampling frequency
$\omega=10$\,\textrm{Hz}, and
the optimal sampling frequency $\omega^*$
that Algorithm~\ref{alg:search} yields.
Note that ETSI~\cite{etsi-va} imposes
$\omega\geq10$\,\textrm{Hz}, however
high precision commercial GPS
devices~\cite{gpscomercial}
offer maximum rates of 10\,\textrm{Hz};
thus setting ETSI positioning
sampling rate to $\omega=10$\,\textrm{Hz}.

Fig.~\ref{fig:optimal_sampling}
numerical results evidence that the optimal
positioning sampling rate $\omega^*$ matches ETSI
sampling rate $\omega=10$\,\textrm{Hz}
in scenarios with
only 16~pedestrians, regardless the
number of cars and bikes.
Hence, ETSI positioning sampling rate is clearly
optimal upon scenarios with low
pedestrian densities.

We now check what happens in scenarios with
32 and 48 pedestrians. If only bikes
are present ---
see~Fig.~\ref{fig:optimal_sampling}~(left) ---
the optimal positioning sampling rate $\omega^*$
always remains below ETSI sampling rate
$\omega=10$\,\textrm{Hz}.
However, upon the presence of cars ---
see~Fig.~\ref{fig:optimal_sampling}~(middle)
and Fig.~\ref{fig:optimal_sampling}~(right) ---
the optimal positioning sampling rate
$\omega^*$ increases with the number of
cars. Given the high CAM rate of
cars $\lambda_c=3$\,\textrm{Hz}, large number
of vehicles lead to more 802.11p
collisions and it
is necessary to send/generate
more pedestrian VAMs (using higher
rates $\omega^*$) to
guarantee their delivery.

However, numerical results highlight
that the optimal sampling frequency
$\omega^*$ increases at a lower pace
upon high density of pedestrians.
For example,
Fig.~\ref{fig:optimal_sampling}~(right)
shows how the optimal sampling rate
$\omega^*$ remains below $10$\,\textrm{Hz}
if 48 pedestrians are in the street,
regardless the number of cars \& bikes.
While the optimal sampling rate
reaches ETSI sampling rate
$\omega=10$\,\textrm{Hz} with just
10 cars \& bikes.

Finally, the numerical results from
Fig.~\ref{fig:optimal_sampling}
evidence how ETSI (markers) results
into worse pIPG than using
the optimal sampling
rate $\omega^*$ obtained
through Algorithm~\ref{alg:search}
(lines).



\subsection{Simulation Results}
\label{subsec:simulation}

In this section we evaluate the
performance of Algorithm~\ref{alg:search}
in simulated scenarios. That is,
rather than using the 802.11p
node model~\cite{BaiocchiAoI}, we
resort to a vehicular simulation stack to
test the performance of the pedestrian VAMs
using the optimal positioning sampling frequencies
found in  \S\ref{subsec:numerical}.

We evaluate the performance of ETSI
positioning sampling frequency
$\omega=10$\,\textrm{Hz} against the
optimal sampling rate $\omega^*$ found
in Algorithm~\ref{alg:search}.
The scenario considered is the same one
from \S\ref{subsec:validation}
and \ref{subsec:numerical}, i.e.
a $2$\,\textrm{km}
street with two sidewalks with pedestrian
and cycle lanes separated by a two-lane road.
In the simulations we consider 48 pedestrians
sharing the 802.11p wireless medium with
0,6,12,24, and 48 cars \& bikes sending
CAMs/VAMs.

The simulation setup we use is Artery~\cite{Artery}, which implements the ETSI ITS protocol stack trough its Vanetza component. Artery uses Veins~\cite{Veins} to implement 802.11p. The movement of pedestrians, cyclists and vehicles is controlled by SUMO~\cite{sumo2012}. We implemented a VRU awareness basic service (VBS) 
following the triggering rules in\cite{etsi-va} and took measurements for 60 seconds after warm-up periods of 100 seconds.
For each simulation we perform five
repetitions.
The simulation parameters are described in Table~\ref{tbl:simpars}.

\begin{table}[t]
	\centering
	\caption{Simulation Parameters}
	\label{tbl:simpars}
	\begin{tabular}{ c  l }
		\toprule
		\textbf{Parameter}  & \textbf{Values} \\
		\midrule
		Access Layer protocol & ITS-G5 (IEEE 802.11p) \\
        Channel bandwidth & 10\,\textrm{MHz} at 5.9\,\textrm{GHz} \\
        Data rate & 6\,\textrm{Mbit/s} \\
        Pedestrian Transmit power & 16\,\textrm{mW}\\
        Bikes/Cars Transmit power & 20\,\textrm{mW}\\
		Path loss model & Two-Ray interference model \\
        VAM generation frequency & 0.2--10~\textrm{Hz} (ETSI VAM~\cite{etsi-va}) \\
        CAM generation frequency & 1--10\,\textrm{Hz} (ETSI CAM~\cite{etsiCAM}) \\
        Max. pedestrian velocity &  5\,\textrm{km/h} \\
        Max. bicycle velocity    & 25\,\textrm{km/h} \\
        Max. vehicle velocity    & 60\,\textrm{km/h} \\
		\bottomrule
	\end{tabular}
\end{table}

\begin{table*}[t]
 \centering
 \caption{Average IGG, iIPG, pIPG and PDR at distances $\leq100$\,\textrm{m} with vaying Pedestrians (P), Bikes (B), and Cars (C).
 Simulations use Artery Veins~\cite{Veins} 802.11p stack,
 and we compare ETSI positioning
 sampling rate~\cite{etsi-va} against
 Algorithm~\ref{alg:search}.
 }
 \begin{tabularx}{\textwidth}{ >{\raggedright\arraybackslash}X |  >{\raggedleft\arraybackslash}X  >{\raggedleft\arraybackslash}X 
 >{\raggedleft\arraybackslash}X  >{\raggedleft\arraybackslash}X  >{\raggedleft\arraybackslash}X | >{\raggedleft\arraybackslash}X 
 >{\raggedleft\arraybackslash}X  >{\raggedleft\arraybackslash}X 
 >{\raggedleft\arraybackslash}X  >{\raggedleft\arraybackslash}X 
 }
 \toprule
 \multirow{2}{*}{\textbf{P-B-C}} & \multicolumn{5}{c|}{\textbf{ETSI}} & \multicolumn{5}{c}{\textbf{Algorithm~\ref{alg:search}}} \\\cline{2-11}
 & \textbf{$\omega$} & \textbf{IGG} & \textbf{iIPG} & \textbf{pIPG} & \textbf{PDR} & \textbf{$\omega$} &\textbf{IGG} &\textbf{iIPG} & \textbf{pIPG} &\textbf{PDR} \\\midrule
 48-0-0 & 10\,\textrm{Hz} & 3.078\,\textrm{sec} & 4.227\,\textrm{sec} & 0.0012\,\textrm{sec} & 0.7282 & 1\,\textrm{Hz} & 3.984\,\textrm{sec} & 4.021\,\textrm{sec} & 0.0046\,\textrm{sec} & 0.9908 \\
 48-6-6 & 10\,\textrm{Hz} & 3.053\,\textrm{sec} & 4.256\,\textrm{sec} & 0.0012\,\textrm{sec} & 0.7174 & 0.8\,\textrm{Hz} & 3.728\,\textrm{sec}  & 3.759\,\textrm{sec} & 0.0058\,\textrm{sec} & 0.9916 \\
 48-12-12 & 10\,\textrm{Hz} & 3.054\,\textrm{sec} &  4.299\,\textrm{sec} & 0.0012\,\textrm{sec} & 0.7104 & 2\,\textrm{Hz} & 3.470\,\textrm{sec}  & 3.496\,\textrm{sec}  & 0.0022\,\textrm{sec} & 0.9925 \\
 48-24-24 & 10\,\textrm{Hz} & 3.055\,\textrm{sec} & 4.461\,\textrm{sec}  & 0.0014\,\textrm{sec} & 0.6826 & 5\,\textrm{Hz} & 3.158\,\textrm{sec} & 3.287\,\textrm{sec}  & 0.0009\,\textrm{sec} & 0.9618 \\
 48-48-48 & 10\,\textrm{Hz} & 3.056\,\textrm{sec} & 4.431\,\textrm{sec}  & 0.0016\,\textrm{sec} & 0.6569 & 7\,\textrm{Hz} & 3.103\,\textrm{sec} & 3.415\,\textrm{sec}  & 0.0007\,\textrm{sec} & 0.9098 \\\bottomrule
 \end{tabularx}
 
 \label{table:ipg_igg_sim}
 \end{table*}

To compare the performance of a VBS
using the positioning sampling rate $\omega^*$ of
Algorithm~\ref{alg:search}, we
consider the following metrics:
\begin{itemize}
    \item \emph{Inter-generation gaps (IGG):} the time between two consecutive pedestrian VAM generations (measured at the transmitting node).
    \item \emph{Pedestrian Inter-packet gaps (pIPG):} the time between the reception of VAMs from \textit{any} pedestrian (measured at receiving nodes --- vehicles and other VRUs).
    \item \emph{Individual Inter-packet gaps (iIPG):} the time between the reception of two VAMs from \textit{one} pedestrian (measured at the receiving nodes).
    \item \emph{Packet Delivery Ratio (PDR):} the ratio between generated messages and successful receptions by neighbors in an area.
\end{itemize}

Inline with \S\ref{subsec:validation},
throughout the simulations
we observe that pedestrians keep their
VAM rate $\lambda$ below $1$\,\textrm{Hz}.
Thus, DCC congestion
control is not triggered for it acts upon
rates exceeding
$\sim3$\,\textrm{Hz}~\cite{Amador2020}.
Consequently, there exists the need of
reducing the congestion with other mechanisms
as the proposed search of an optimal
positioning sampling rate $\omega^*$.


\begin{figure*}[t]
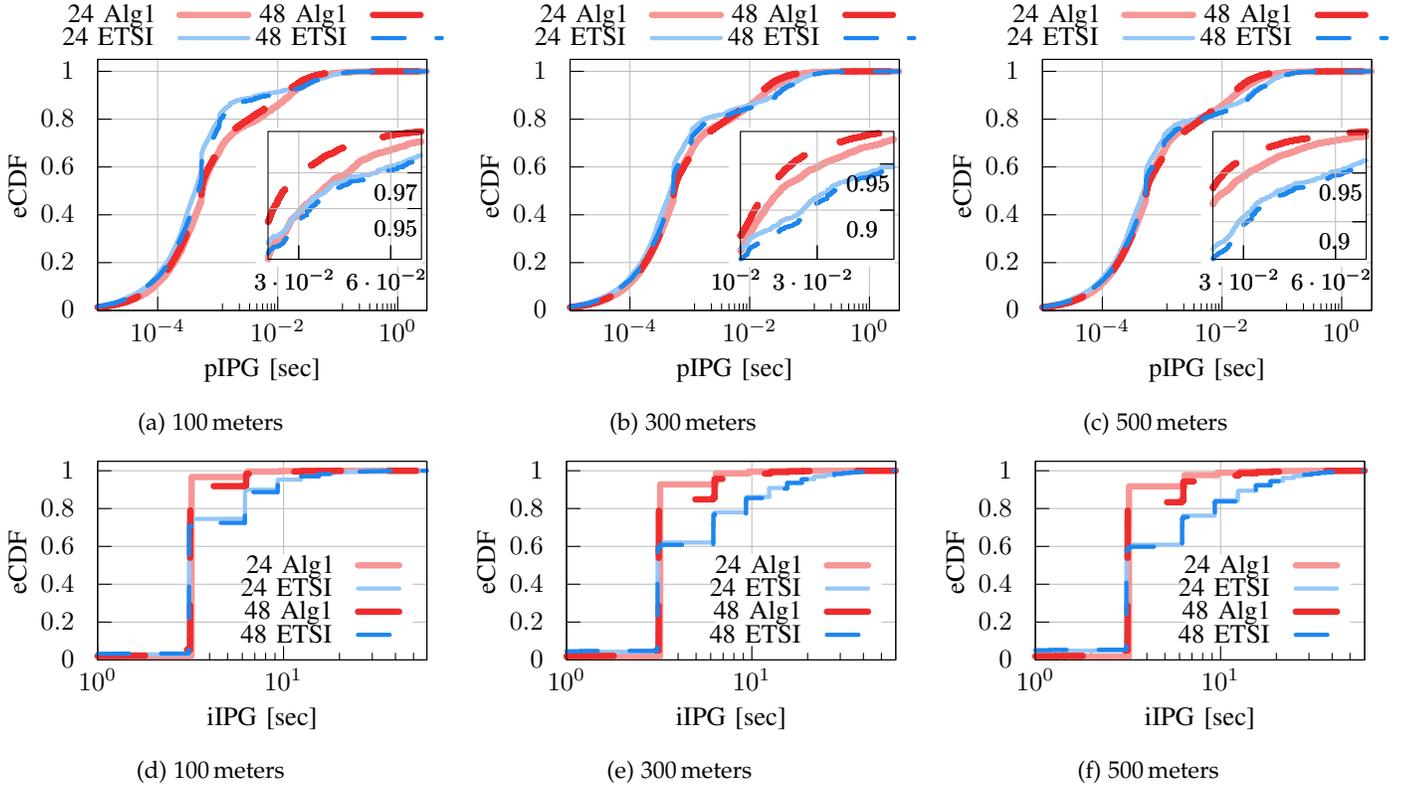

    \centering

    \begin{subfigure}{0.31\textwidth}
        \centering
        \input{fig/pipg100mlogcdf.tex}
        \vspace{-1em}
        \caption{100\,\textrm{meters}}
        \label{fig:pipg100mcdf}
    \end{subfigure}
    \hfill
    \begin{subfigure}{0.31\textwidth}
        \centering
        \input{fig/pipg300mlogcdf.tex}
        \vspace{-1em}
        \caption{300\,\textrm{meters}}
        \label{fig:pipg300mcdf}
    \end{subfigure}
    \hfill
    \begin{subfigure}{0.31\textwidth}
        \centering
        \input{fig/pipg500mlogcdf.tex}
        \vspace{-1em}
        \caption{500\,\textrm{meters}}
        \label{fig:pipg500mcdf}
    \end{subfigure}
    \hfill
    \\
    \begin{subfigure}{0.31\textwidth}
        \centering
        \input{fig/100mlogcdf.tex}
        \vspace{-1em}
        \caption{100\,\textrm{meters}}
        \label{fig:100mcdf}
    \end{subfigure}
    \hfill
    \begin{subfigure}{0.31\textwidth}
        \centering
        \input{fig/300mlogcdf.tex}
        \vspace{-1em}
        \caption{300\,\textrm{meters}}
        \label{fig:300mcdf}
    \end{subfigure}
    \hfill
    \begin{subfigure}{0.31\textwidth}
        \centering
        \input{fig/500mlogcdf}
        \vspace{-1em}
        \caption{500\,\textrm{meters}}
        \label{fig:500mcdf}
    \end{subfigure}
    \hfill
    \caption{pIPG (top) and iIPG (down)
    empirical CDF (eCDF)
    for pedestrians at different distances
    using
    Artery Veins~\cite{Veins} 802.11p stack.
    We compare
    ETSI positioning
    sampling rate~\cite{etsi-va}
    (blue) against
    the Algorithm~\ref{alg:search} (red) in scenarios
    with:
    48 pedestrians and 24 bikes \&
    vehicles (continuous), and
    48 pedestrians with 48 bikes \&
    vehicles (dashed). Optimal sampling
    rates are $\omega=5$\,\textrm{Hz}
    and $\omega=7$\,\textrm{Hz}
    for 24 and 48 bikes \& vehicles,
    respectively.}
    \label{fig:cdf_iipg_distance}
\end{figure*}

Table~\ref{table:ipg_igg_sim} shows the average IGGs, iIPGs, pIPG, and PDRs for different scenarios, starting from 48 pedestrians and no bicycles or vehicles (P-B-V 48-0-0), up to scenarios with 24 bicycles and 24 vehicles (P-B-V 48-24-24). We compare the iIPG achieved by the ETSI parameters ($\omega = 10$\,Hz) and the
positioning sampling obtained through
Algorithm~\ref{alg:search}
-- see~Fig.~\ref{fig:optimal_sampling}.
Again, note that ETSI~\cite{etsi-va}
imposes $\omega\geq10$\,\textrm{Hz}, but
high precision commercial GPS
devices~\cite{gpscomercial} offer maximum
rates of 10\,\textrm{Hz}. Thus, we
choose ETSI positioning sampling rate
$\omega=10$\,\textrm{Hz}.

Furthermore, it is shown from the get-go that the optimal values for $\omega$ outperform the ETSI scheme significantly, since neighbors hear from pedestrians between 0.2 and 1.17\,s more frequently. It is possible to compare the intended update rate from pedestrians (IGG) and the actual result in iIPG. The table shows that optimal values for $\omega$ take advantage of the space left by bicycles and vehicles more efficiently, i.e., practically all of the information intended to reach neighboring nodes is received, while the ETSI schemes loses around 30\% of the messages. 

For the scenarios with fewer bicycles and cars (48-0-0, 48-6-6), ETSI achieves lower values for pIPG than Algorithm~\ref{alg:search}. This is due to the fact that assumptions \ref{assumption:1} and \ref{assumption:2} do not hold, and the objective PDR for Algorithm~\ref{alg:search} is not correct. However, in scenarios with more bicycles and cars (48-24-44, 48-48-46), Palm-Khintchine's theorem~\cite{p-k} hold, which is reflected in the PDR input for Algorithm~\ref{alg:search} and thus average pIPG is significantly lower. Nevertheless, even if ETSI has better average pIPG in some scenarios, its values for iIPG and PDR are always outperformed by Algorithm~\ref{alg:search}. The lower average values for iIPG mean that pedestrians using Algorithm~\ref{alg:search} are tracked better and more often than those using the ETSI mechanism. 

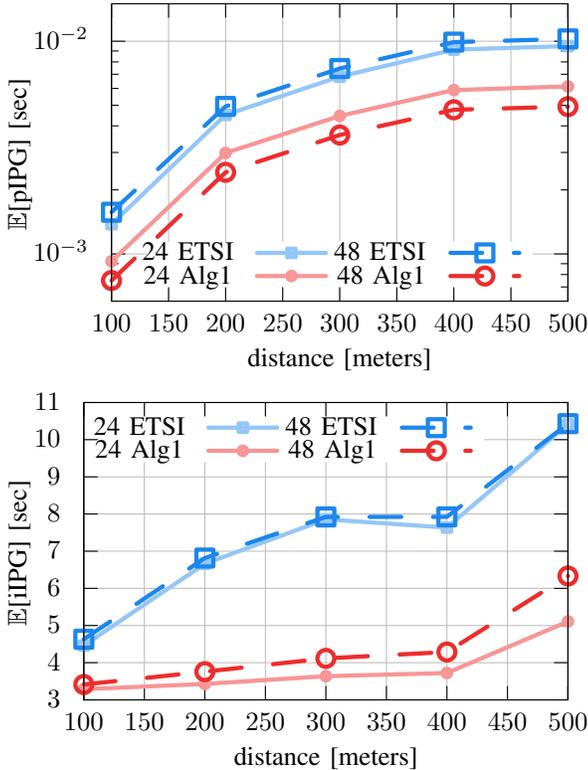
\begin{figure}[t]
    \centering
    \begin{subfigure}{\textwidth}
        \begin{tikzpicture}[gnuplot]
\path (0.000,0.000) rectangle (8.125,5.250);
\gpcolor{color=gp lt color border}
\gpsetlinetype{gp lt border}
\gpsetdashtype{gp dt solid}
\gpsetlinewidth{1.00}
\draw[gp path] (1.504,0.985)--(1.594,0.985);
\draw[gp path] (7.572,0.985)--(7.482,0.985);
\draw[gp path] (1.504,1.174)--(1.594,1.174);
\draw[gp path] (7.572,1.174)--(7.482,1.174);
\draw[gp path] (1.504,1.339)--(1.594,1.339);
\draw[gp path] (7.572,1.339)--(7.482,1.339);
\draw[gp path] (1.504,1.483)--(1.594,1.483);
\draw[gp path] (7.572,1.483)--(7.482,1.483);
\gpcolor{rgb color={0.745,0.745,0.745}}
\draw[gp path] (1.504,1.613)--(1.966,1.613);
\draw[gp path] (7.110,1.613)--(7.572,1.613);
\gpcolor{color=gp lt color border}
\draw[gp path] (1.504,1.613)--(1.684,1.613);
\draw[gp path] (7.572,1.613)--(7.392,1.613);
\node[gp node right] at (1.320,1.613) {$10^{-3}$};
\draw[gp path] (1.504,2.465)--(1.594,2.465);
\draw[gp path] (7.572,2.465)--(7.482,2.465);
\draw[gp path] (1.504,2.963)--(1.594,2.963);
\draw[gp path] (7.572,2.963)--(7.482,2.963);
\draw[gp path] (1.504,3.317)--(1.594,3.317);
\draw[gp path] (7.572,3.317)--(7.482,3.317);
\draw[gp path] (1.504,3.591)--(1.594,3.591);
\draw[gp path] (7.572,3.591)--(7.482,3.591);
\draw[gp path] (1.504,3.815)--(1.594,3.815);
\draw[gp path] (7.572,3.815)--(7.482,3.815);
\draw[gp path] (1.504,4.004)--(1.594,4.004);
\draw[gp path] (7.572,4.004)--(7.482,4.004);
\draw[gp path] (1.504,4.168)--(1.594,4.168);
\draw[gp path] (7.572,4.168)--(7.482,4.168);
\draw[gp path] (1.504,4.313)--(1.594,4.313);
\draw[gp path] (7.572,4.313)--(7.482,4.313);
\gpcolor{rgb color={0.745,0.745,0.745}}
\draw[gp path] (1.504,4.443)--(7.572,4.443);
\gpcolor{color=gp lt color border}
\draw[gp path] (1.504,4.443)--(1.684,4.443);
\draw[gp path] (7.572,4.443)--(7.392,4.443);
\node[gp node right] at (1.320,4.443) {$10^{-2}$};
\gpcolor{rgb color={0.745,0.745,0.745}}
\draw[gp path] (1.504,0.985)--(1.504,4.941);
\gpcolor{color=gp lt color border}
\draw[gp path] (1.504,0.985)--(1.504,1.165);
\draw[gp path] (1.504,4.941)--(1.504,4.761);
\node[gp node center] at (1.504,0.677) {$100$};
\gpcolor{rgb color={0.745,0.745,0.745}}
\draw[gp path] (2.263,0.985)--(2.263,1.165);
\draw[gp path] (2.263,1.781)--(2.263,4.941);
\gpcolor{color=gp lt color border}
\draw[gp path] (2.263,0.985)--(2.263,1.165);
\draw[gp path] (2.263,4.941)--(2.263,4.761);
\node[gp node center] at (2.263,0.677) {$150$};
\gpcolor{rgb color={0.745,0.745,0.745}}
\draw[gp path] (3.021,0.985)--(3.021,1.165);
\draw[gp path] (3.021,1.781)--(3.021,4.941);
\gpcolor{color=gp lt color border}
\draw[gp path] (3.021,0.985)--(3.021,1.165);
\draw[gp path] (3.021,4.941)--(3.021,4.761);
\node[gp node center] at (3.021,0.677) {$200$};
\gpcolor{rgb color={0.745,0.745,0.745}}
\draw[gp path] (3.780,0.985)--(3.780,1.165);
\draw[gp path] (3.780,1.781)--(3.780,4.941);
\gpcolor{color=gp lt color border}
\draw[gp path] (3.780,0.985)--(3.780,1.165);
\draw[gp path] (3.780,4.941)--(3.780,4.761);
\node[gp node center] at (3.780,0.677) {$250$};
\gpcolor{rgb color={0.745,0.745,0.745}}
\draw[gp path] (4.538,0.985)--(4.538,1.165);
\draw[gp path] (4.538,1.781)--(4.538,4.941);
\gpcolor{color=gp lt color border}
\draw[gp path] (4.538,0.985)--(4.538,1.165);
\draw[gp path] (4.538,4.941)--(4.538,4.761);
\node[gp node center] at (4.538,0.677) {$300$};
\gpcolor{rgb color={0.745,0.745,0.745}}
\draw[gp path] (5.297,0.985)--(5.297,1.165);
\draw[gp path] (5.297,1.781)--(5.297,4.941);
\gpcolor{color=gp lt color border}
\draw[gp path] (5.297,0.985)--(5.297,1.165);
\draw[gp path] (5.297,4.941)--(5.297,4.761);
\node[gp node center] at (5.297,0.677) {$350$};
\gpcolor{rgb color={0.745,0.745,0.745}}
\draw[gp path] (6.055,0.985)--(6.055,1.165);
\draw[gp path] (6.055,1.781)--(6.055,4.941);
\gpcolor{color=gp lt color border}
\draw[gp path] (6.055,0.985)--(6.055,1.165);
\draw[gp path] (6.055,4.941)--(6.055,4.761);
\node[gp node center] at (6.055,0.677) {$400$};
\gpcolor{rgb color={0.745,0.745,0.745}}
\draw[gp path] (6.814,0.985)--(6.814,1.165);
\draw[gp path] (6.814,1.781)--(6.814,4.941);
\gpcolor{color=gp lt color border}
\draw[gp path] (6.814,0.985)--(6.814,1.165);
\draw[gp path] (6.814,4.941)--(6.814,4.761);
\node[gp node center] at (6.814,0.677) {$450$};
\gpcolor{rgb color={0.745,0.745,0.745}}
\draw[gp path] (7.572,0.985)--(7.572,4.941);
\gpcolor{color=gp lt color border}
\draw[gp path] (7.572,0.985)--(7.572,1.165);
\draw[gp path] (7.572,4.941)--(7.572,4.761);
\node[gp node center] at (7.572,0.677) {$500$};
\draw[gp path] (1.504,4.941)--(1.504,0.985)--(7.572,0.985)--(7.572,4.941)--cycle;
\node[gp node center,rotate=-270] at (0.292,2.963) {$\mathbb{E}$[pIPG] [sec]};
\node[gp node center] at (4.538,0.215) {distance [meters]};
\node[gp node right] at (3.254,1.627) {24 ETSI};
\gpcolor{rgb color={0.580,0.776,0.969}}
\gpsetlinewidth{4.00}
\draw[gp path] (3.438,1.627)--(4.354,1.627);
\draw[gp path] (1.504,2.008)--(3.021,3.463)--(4.538,3.972)--(6.055,4.331)--(7.572,4.378);
\gpsetpointsize{4.00}
\gppoint{gp mark 4}{(1.504,2.008)}
\gppoint{gp mark 4}{(3.021,3.463)}
\gppoint{gp mark 4}{(4.538,3.972)}
\gppoint{gp mark 4}{(6.055,4.331)}
\gppoint{gp mark 4}{(7.572,4.378)}
\gppoint{gp mark 4}{(3.896,1.627)}
\gpcolor{color=gp lt color border}
\node[gp node right] at (3.254,1.319) {24 Alg1};
\gpcolor{rgb color={0.969,0.588,0.588}}
\draw[gp path] (3.438,1.319)--(4.354,1.319);
\draw[gp path] (1.504,1.515)--(3.021,2.956)--(4.538,3.449)--(6.055,3.793)--(7.572,3.842);
\gppoint{gp mark 6}{(1.504,1.515)}
\gppoint{gp mark 6}{(3.021,2.956)}
\gppoint{gp mark 6}{(4.538,3.449)}
\gppoint{gp mark 6}{(6.055,3.793)}
\gppoint{gp mark 6}{(7.572,3.842)}
\gppoint{gp mark 6}{(3.896,1.319)}
\gpcolor{color=gp lt color border}
\node[gp node right] at (5.826,1.627) {48 ETSI};
\gpcolor{rgb color={0.110,0.525,0.933}}
\gpsetdashtype{gp dt 2}
\draw[gp path] (6.010,1.627)--(6.926,1.627);
\draw[gp path] (1.504,2.168)--(3.021,3.576)--(4.538,4.080)--(6.055,4.431)--(7.572,4.474);
\gpsetpointsize{8.00}
\gppoint{gp mark 4}{(1.504,2.168)}
\gppoint{gp mark 4}{(3.021,3.576)}
\gppoint{gp mark 4}{(4.538,4.080)}
\gppoint{gp mark 4}{(6.055,4.431)}
\gppoint{gp mark 4}{(7.572,4.474)}
\gppoint{gp mark 4}{(6.468,1.627)}
\gpcolor{color=gp lt color border}
\node[gp node right] at (5.826,1.319) {48 Alg1};
\gpcolor{rgb color={0.933,0.173,0.173}}
\draw[gp path] (6.010,1.319)--(6.926,1.319);
\draw[gp path] (1.504,1.260)--(3.021,2.699)--(4.538,3.195)--(6.055,3.531)--(7.572,3.573);
\gppoint{gp mark 6}{(1.504,1.260)}
\gppoint{gp mark 6}{(3.021,2.699)}
\gppoint{gp mark 6}{(4.538,3.195)}
\gppoint{gp mark 6}{(6.055,3.531)}
\gppoint{gp mark 6}{(7.572,3.573)}
\gppoint{gp mark 6}{(6.468,1.319)}
\gpcolor{color=gp lt color border}
\gpsetdashtype{gp dt solid}
\gpsetlinewidth{1.00}
\draw[gp path] (1.504,4.941)--(1.504,0.985)--(7.572,0.985)--(7.572,4.941)--cycle;
\gpdefrectangularnode{gp plot 1}{\pgfpoint{1.504cm}{0.985cm}}{\pgfpoint{7.572cm}{4.941cm}}
\end{tikzpicture}
    \end{subfigure}\\
    \begin{subfigure}{\textwidth}
        \begin{tikzpicture}[gnuplot]
\path (0.000,0.000) rectangle (8.125,5.250);
\gpcolor{rgb color={0.745,0.745,0.745}}
\gpsetlinetype{gp lt border}
\gpsetdashtype{gp dt solid}
\gpsetlinewidth{1.00}
\draw[gp path] (1.136,0.985)--(7.572,0.985);
\gpcolor{color=gp lt color border}
\draw[gp path] (1.136,0.985)--(1.316,0.985);
\draw[gp path] (7.572,0.985)--(7.392,0.985);
\node[gp node right] at (0.952,0.985) {$3$};
\gpcolor{rgb color={0.745,0.745,0.745}}
\draw[gp path] (1.136,1.480)--(7.572,1.480);
\gpcolor{color=gp lt color border}
\draw[gp path] (1.136,1.480)--(1.316,1.480);
\draw[gp path] (7.572,1.480)--(7.392,1.480);
\node[gp node right] at (0.952,1.480) {$4$};
\gpcolor{rgb color={0.745,0.745,0.745}}
\draw[gp path] (1.136,1.974)--(7.572,1.974);
\gpcolor{color=gp lt color border}
\draw[gp path] (1.136,1.974)--(1.316,1.974);
\draw[gp path] (7.572,1.974)--(7.392,1.974);
\node[gp node right] at (0.952,1.974) {$5$};
\gpcolor{rgb color={0.745,0.745,0.745}}
\draw[gp path] (1.136,2.469)--(7.572,2.469);
\gpcolor{color=gp lt color border}
\draw[gp path] (1.136,2.469)--(1.316,2.469);
\draw[gp path] (7.572,2.469)--(7.392,2.469);
\node[gp node right] at (0.952,2.469) {$6$};
\gpcolor{rgb color={0.745,0.745,0.745}}
\draw[gp path] (1.136,2.963)--(7.572,2.963);
\gpcolor{color=gp lt color border}
\draw[gp path] (1.136,2.963)--(1.316,2.963);
\draw[gp path] (7.572,2.963)--(7.392,2.963);
\node[gp node right] at (0.952,2.963) {$7$};
\gpcolor{rgb color={0.745,0.745,0.745}}
\draw[gp path] (1.136,3.458)--(7.572,3.458);
\gpcolor{color=gp lt color border}
\draw[gp path] (1.136,3.458)--(1.316,3.458);
\draw[gp path] (7.572,3.458)--(7.392,3.458);
\node[gp node right] at (0.952,3.458) {$8$};
\gpcolor{rgb color={0.745,0.745,0.745}}
\draw[gp path] (1.136,3.952)--(7.572,3.952);
\gpcolor{color=gp lt color border}
\draw[gp path] (1.136,3.952)--(1.316,3.952);
\draw[gp path] (7.572,3.952)--(7.392,3.952);
\node[gp node right] at (0.952,3.952) {$9$};
\gpcolor{rgb color={0.745,0.745,0.745}}
\draw[gp path] (1.136,4.447)--(1.320,4.447);
\draw[gp path] (6.464,4.447)--(7.572,4.447);
\gpcolor{color=gp lt color border}
\draw[gp path] (1.136,4.447)--(1.316,4.447);
\draw[gp path] (7.572,4.447)--(7.392,4.447);
\node[gp node right] at (0.952,4.447) {$10$};
\gpcolor{rgb color={0.745,0.745,0.745}}
\draw[gp path] (1.136,4.941)--(7.572,4.941);
\gpcolor{color=gp lt color border}
\draw[gp path] (1.136,4.941)--(1.316,4.941);
\draw[gp path] (7.572,4.941)--(7.392,4.941);
\node[gp node right] at (0.952,4.941) {$11$};
\gpcolor{rgb color={0.745,0.745,0.745}}
\draw[gp path] (1.136,0.985)--(1.136,4.941);
\gpcolor{color=gp lt color border}
\draw[gp path] (1.136,0.985)--(1.136,1.165);
\draw[gp path] (1.136,4.941)--(1.136,4.761);
\node[gp node center] at (1.136,0.677) {$100$};
\gpcolor{rgb color={0.745,0.745,0.745}}
\draw[gp path] (1.941,0.985)--(1.941,4.145);
\draw[gp path] (1.941,4.761)--(1.941,4.941);
\gpcolor{color=gp lt color border}
\draw[gp path] (1.941,0.985)--(1.941,1.165);
\draw[gp path] (1.941,4.941)--(1.941,4.761);
\node[gp node center] at (1.941,0.677) {$150$};
\gpcolor{rgb color={0.745,0.745,0.745}}
\draw[gp path] (2.745,0.985)--(2.745,4.145);
\draw[gp path] (2.745,4.761)--(2.745,4.941);
\gpcolor{color=gp lt color border}
\draw[gp path] (2.745,0.985)--(2.745,1.165);
\draw[gp path] (2.745,4.941)--(2.745,4.761);
\node[gp node center] at (2.745,0.677) {$200$};
\gpcolor{rgb color={0.745,0.745,0.745}}
\draw[gp path] (3.550,0.985)--(3.550,4.145);
\draw[gp path] (3.550,4.761)--(3.550,4.941);
\gpcolor{color=gp lt color border}
\draw[gp path] (3.550,0.985)--(3.550,1.165);
\draw[gp path] (3.550,4.941)--(3.550,4.761);
\node[gp node center] at (3.550,0.677) {$250$};
\gpcolor{rgb color={0.745,0.745,0.745}}
\draw[gp path] (4.354,0.985)--(4.354,4.145);
\draw[gp path] (4.354,4.761)--(4.354,4.941);
\gpcolor{color=gp lt color border}
\draw[gp path] (4.354,0.985)--(4.354,1.165);
\draw[gp path] (4.354,4.941)--(4.354,4.761);
\node[gp node center] at (4.354,0.677) {$300$};
\gpcolor{rgb color={0.745,0.745,0.745}}
\draw[gp path] (5.159,0.985)--(5.159,4.145);
\draw[gp path] (5.159,4.761)--(5.159,4.941);
\gpcolor{color=gp lt color border}
\draw[gp path] (5.159,0.985)--(5.159,1.165);
\draw[gp path] (5.159,4.941)--(5.159,4.761);
\node[gp node center] at (5.159,0.677) {$350$};
\gpcolor{rgb color={0.745,0.745,0.745}}
\draw[gp path] (5.963,0.985)--(5.963,4.145);
\draw[gp path] (5.963,4.761)--(5.963,4.941);
\gpcolor{color=gp lt color border}
\draw[gp path] (5.963,0.985)--(5.963,1.165);
\draw[gp path] (5.963,4.941)--(5.963,4.761);
\node[gp node center] at (5.963,0.677) {$400$};
\gpcolor{rgb color={0.745,0.745,0.745}}
\draw[gp path] (6.768,0.985)--(6.768,4.941);
\gpcolor{color=gp lt color border}
\draw[gp path] (6.768,0.985)--(6.768,1.165);
\draw[gp path] (6.768,4.941)--(6.768,4.761);
\node[gp node center] at (6.768,0.677) {$450$};
\gpcolor{rgb color={0.745,0.745,0.745}}
\draw[gp path] (7.572,0.985)--(7.572,4.941);
\gpcolor{color=gp lt color border}
\draw[gp path] (7.572,0.985)--(7.572,1.165);
\draw[gp path] (7.572,4.941)--(7.572,4.761);
\node[gp node center] at (7.572,0.677) {$500$};
\draw[gp path] (1.136,4.941)--(1.136,0.985)--(7.572,0.985)--(7.572,4.941)--cycle;
\node[gp node center,rotate=-270] at (0.292,2.963) {$\mathbb{E}$[iIPG] [sec]};
\node[gp node center] at (4.354,0.215) {distance [meters]};
\node[gp node right] at (2.608,4.607) {24 ETSI};
\gpcolor{rgb color={0.580,0.776,0.969}}
\gpsetlinewidth{4.00}
\draw[gp path] (2.792,4.607)--(3.708,4.607);
\draw[gp path] (1.136,1.708)--(2.745,2.796)--(4.354,3.387)--(5.963,3.277)--(7.572,4.652);
\gpsetpointsize{4.00}
\gppoint{gp mark 4}{(1.136,1.708)}
\gppoint{gp mark 4}{(2.745,2.796)}
\gppoint{gp mark 4}{(4.354,3.387)}
\gppoint{gp mark 4}{(5.963,3.277)}
\gppoint{gp mark 4}{(7.572,4.652)}
\gppoint{gp mark 4}{(3.250,4.607)}
\gpcolor{color=gp lt color border}
\node[gp node right] at (2.608,4.299) {24 Alg1};
\gpcolor{rgb color={0.969,0.588,0.588}}
\draw[gp path] (2.792,4.299)--(3.708,4.299);
\draw[gp path] (1.136,1.127)--(2.745,1.197)--(4.354,1.300)--(5.963,1.342)--(7.572,2.028);
\gppoint{gp mark 6}{(1.136,1.127)}
\gppoint{gp mark 6}{(2.745,1.197)}
\gppoint{gp mark 6}{(4.354,1.300)}
\gppoint{gp mark 6}{(5.963,1.342)}
\gppoint{gp mark 6}{(7.572,2.028)}
\gppoint{gp mark 6}{(3.250,4.299)}
\gpcolor{color=gp lt color border}
\node[gp node right] at (5.180,4.607) {48 ETSI};
\gpcolor{rgb color={0.110,0.525,0.933}}
\gpsetdashtype{gp dt 2}
\draw[gp path] (5.364,4.607)--(6.280,4.607);
\draw[gp path] (1.136,1.792)--(2.745,2.872)--(4.354,3.419)--(5.963,3.420)--(7.572,4.660);
\gpsetpointsize{8.00}
\gppoint{gp mark 4}{(1.136,1.792)}
\gppoint{gp mark 4}{(2.745,2.872)}
\gppoint{gp mark 4}{(4.354,3.419)}
\gppoint{gp mark 4}{(5.963,3.420)}
\gppoint{gp mark 4}{(7.572,4.660)}
\gppoint{gp mark 4}{(5.822,4.607)}
\gpcolor{color=gp lt color border}
\node[gp node right] at (5.180,4.299) {48 Alg1};
\gpcolor{rgb color={0.933,0.173,0.173}}
\draw[gp path] (5.364,4.299)--(6.280,4.299);
\draw[gp path] (1.136,1.190)--(2.745,1.360)--(4.354,1.537)--(5.963,1.622)--(7.572,2.633);
\gppoint{gp mark 6}{(1.136,1.190)}
\gppoint{gp mark 6}{(2.745,1.360)}
\gppoint{gp mark 6}{(4.354,1.537)}
\gppoint{gp mark 6}{(5.963,1.622)}
\gppoint{gp mark 6}{(7.572,2.633)}
\gppoint{gp mark 6}{(5.822,4.299)}
\gpcolor{color=gp lt color border}
\gpsetdashtype{gp dt solid}
\gpsetlinewidth{1.00}
\draw[gp path] (1.136,4.941)--(1.136,0.985)--(7.572,0.985)--(7.572,4.941)--cycle;
\gpdefrectangularnode{gp plot 1}{\pgfpoint{1.136cm}{0.985cm}}{\pgfpoint{7.572cm}{4.941cm}}
\end{tikzpicture}
    \end{subfigure}
    \caption{Average
    pIPG (top) and
    iIPG (bottom) of
    pedestrians at different distances
    using
    Artery Veins~\cite{Veins} 802.11p stack.
    We compare
    ETSI positioning
    sampling rate~\cite{etsi-va}
    (blue) against
    the Algorithm~\ref{alg:search} (red) in scenarios
    with:
    48 pedestrians and 24 bikes \&
    vehicles (continuous), and
    48 pedestrians with 48 bikes \&
    vehicles (dashed). Optimal sampling
    rates are $\omega=5$\,\textrm{Hz}
    and $\omega=7$\,\textrm{Hz}
    for 24 and 48 bikes \& vehicles,
    respectively.
    }
    \label{fig:avg_iipg_distance}
\end{figure}


The PDR values from Table~\ref{table:ipg_igg_sim} have implications in other performance metrics such as energy consumption. Every message that is transmitted consumes a certain amount of energy which can be obtained by multiplying the transmission power (16~mW) and the time it takes for a transmission to occur (i.e., 467.424 $\mu$s for a VAM). Schemes using the optimal $\omega$ transmit between 3 and 30\% fewer messages than the ETSI scheme, which then wastes around 30\% of its transmission power consumption in colliding messages. All in all, optimal $\omega$ schemes is more efficient in resource usage, from energy to medium.


Fig.~\ref{fig:cdf_iipg_distance} shows the effect of $\omega$ on pIPG and iIPG. The upper part of the figure shows pIPG results for ETSI VAM and Algorithm~\ref{alg:search}. While lines for ETSI and Algorithm~\ref{alg:search} overlap, a zoom into important regions is also provided. There, it is noticeable that pIPG values for Algorithm~\ref{alg:search} stay lower than those for ETSI and this difference increases with distance. The results for ETSI show that extremely large values for pIPG occur in all distances (i.e., its distribution has a longer tail).

Moreover, the lower part of Fig.~\ref{fig:cdf_iipg_distance} shows results for iIPG. Here, results show that Algorithm~\ref{alg:search} has better iIPGs in all percentiles, while the ETSI mechanism has a significant number of neighbors waiting more than 10 seconds between updates. Scenarios using the calculated optimal $\omega$ are more reliable, and deliver most updates at rates closer to the IGGs from Table~\ref{table:ipg_igg_sim}. While pIPG is a measure of efficient channel utilization, iIPG can be translated to a safety metric: \textit{awareness}. In safety terms, Algorithm 1 outperforms ETSI VAM significantly by allowing nodes keep better track of pedestrians by minimizing the time between two successful updates.

Fig.~\ref{fig:avg_iipg_distance} shows results for pIPG and iIPG over the distance. Once again, in crowded scenarios, Algorithm 1 manages to minimize pIPG. Regarding awareness, iIPG values stay stable through longer distances when optimal values for $\omega$ are used. Receptions at longer distances are affected by attenuation and phenomena such as hidden nodes. However, Akgorithm 1 manages to stay stable up to 400\,m. These results validate the analysis shown in \S\ref{sec:results}, and emphasizes the importance of the optimal sampling rate to increase VRU safety.

Results from the experimental evaluation of the effect of optimal sampling rates for VAMs confirm what the analytical results proposed. First, given that the dynamics of a pedestrian are different to those of a vehicle, positioning sampling can be less aggressive and still react to changes in speed and orientation. Second, that even if less aggressive schemes are used and generation frequencies are lowered, messages are received more frequently with less greedy sampling rates. Finally, optimal values for $\omega$ affect performance favorably in the most important metric: awareness. Other stations are able to see pedestrians more frequently and at longer distances, a significant improvement from the standardized scheme.

\section{Related Work}
\label{sec:related}

The dissemination of awareness messages,
e.g., VAMs and CAMs,
is one of the main topics in the study of V2X and P2X
communications. Additionally, it is necessary
to comprehend the wireless channel access
to analyze the performance of awareness
services. Existing work in the literature
has analyzed the aforementioned topics,
and their results relate to the research
of the present manuscript.

\emph{Wireless Channel Access}.
Cooperative awareness in V2X/P2X communications
use Direct Short Range Communication
technologies (DSRC) as 802.11p/bd, and
C-V2X to deliver VAMs/CAMs. 
The literature has already analyzed the
differences between both C-V2X and
DSRC technologies through simulation
in the physical
layer~\cite{compare-80211-cell},
and field tests~\cite{dsrc-vs-lte}.
Moreover, the research community has also
studied the access layer of both DSRC
and C-V2X technologies. In particular,
works as~\cite{sidelink-persistence}
study how persistent scheduling impacts
the Age of Information (AoI)
in the NR sidelink,
while others as~\cite{BaiocchiAoI}
focus on the impact of 802.11p congestion
on the AoI when DCC does not apply
(as in the case of pedestrian VAMs).
Moreover, works as~\cite{OscarGOT} propose
mechanisms to improve the DCC efficiency,
and assess its performance through
simulation~\cite{Amador2020}.

\emph{CAMs}. Given the rise of autonomous
vehicles, the literature has also 
studied recently the performance of CAMs.
Namely, works as~\cite{cam-generation-model}
model how CAMs are generated depending on
kinematic triggers (e.g. change of position),
while other works focus on how performance
evaluation on either NR sidelink
channels~\cite{cam-sidelink-reliability} and
802.11p channels~\cite{Lyamin2017}.
On top, researchers have also come up with
mechanisms to: adapt the CAM generation
in windy scenarios~\cite{cam-wind};
anticipate to dangerous events through
Machine Learning~\cite{ml-cams}; or
combine sensing information from vehicles,
pedestrians and infrastructure to enhance
vehicle awareness~\cite{combine}.

\emph{VRUs and VAMs}.
Works as~\cite{Bruno2023} state
potential of sending
awareness messages concerning
pedestrians/VRUs in future mobility
scenarios -- i.e. VAMs.
Either by assessing its performance
with onboard units~\cite{cv2xmobilecase}
or roadside cameras~\cite{IslamPSM},
results show the potential of VAMs to
prevent accidents.
In particular, the research community
has assessed the awareness performance
of ETSI VA~\cite{etsi-va}
in LTE~\cite{VAMv2xField},
802.11p/bd and C-V2X PC5~\cite{assessvru}.
Moreover, the literature has also studied
how to increase the awareness level
of ETSI VAMs using passive perception
-- e.g. vehicle camera detecting a pedestrian
-- together with the active transmission
of VAMs~\cite{combine,Teixeira2023}.
Furthermore, works as~\cite{cluster-eval}
study how VRU clusters decrease the collision
of VAMs in the wireless channel upon
high density of pedestrians.

Despite the existing literature on VAM/CAMs,
to the best of our knowledge
there is no analysis
on how decreasing the positioning
sampling rate may enhance the awareness
of ETSI VAMs coming from pedestrians.
Our work aims to fill such a gap
($i$) with a detailed characterization
of the pedestrian VAMs through the
well-known GNM mobility
model~\cite{pedestrians}; and
($ii$) an optimization problem that minimizes
the time between pedestrian VAMs through
adequate election of the positioning
sampling rate.

\section{Conclusion and future work}
In this work we study how to improve 
the delivery of pedestrian VAMs through
optimal positioning sampling. We characterize
how the positioning sampling rate impacts
the rate at which VAMs are generated
when pedestrians change their: position,
speed, and orientation. Simulations
prove the validity of our characterization,
which we leverage to design an algorithm
that finds the optimal
positioning sampling rate in 802.11p channels.
Results show that our algorithm finds
smaller positioning sampling rates than the
proposed by ETSI ($\omega=10$\,\textrm{Hz})
to check if a VAM must be generated.
Moreover, simulations with 24 and 48
cars \& bikes evidence that
decreasing the positioning sampling rate by
a $50\%$ and $30\%$, respectively,
results into reducing
the inter packet gap of pedestrian VAMs
and an increasing the VAM
delivery ratio.
Hence, the optimal positioning sampling rates
found by our algorithm result into
battery savings and higher
pedestrian safety.

In future work we plan to extend our
simulation campaign to even denser
scenarios, and
investigate how the optimal positioning sampling rate
varies using other wireless technologies
as the NR sidelink or 802.11bd.
Additionally, we aim to study:
($i$) other VRU profiles
(e.g. bikes and
scooters);
($ii$) the impact of
the pedestrians'
height and their
ongoing activity
(e.g. listening
to music) on the
VAM rate;
($iii$) whether intention sharing
alters the VAM rate;
and ($iv$) how VRU
clustering may allow
reducing the positioning
sampling rate, yet
keeping the
VAM delivery ratio.



\bibliographystyle{IEEEtran}
\bibliography{mybibfile}

\appendices

\section{Bounding the
Orientation Change}
\label{app:bound}

To bound the orientation
change
$\theta(i\omega^{-1})$
in a sampling period we
cannot resort to closed
form solutions of the
ODE system because it is
not a constant coefficient
system, nor a time dependant
coefficient
system -- see~\cite{ode}.
That is, we cannot express
\eqref{eq:x-der}-\eqref{eq:w-der}
as a system
$\dot{y}(t)=Ay(t)$ nor
$\dot{y}(t)=A(t)y(t)$.
Consequently we cannot use
exponentiation to solve the
ODE system.

To bound the orientation
we take as reference
the vector $(1,0)$
as we did in  \S\ref{sec:orient}.
Hence the second coordinate
is fully
determined by the former
and the change of
orientation, i.e. 
$x_2(t)=x_1(t)\sin(\theta(t))$
with $x(t)=(x_1(t),x_2(t))$.
Therfore, \eqref{eq:orient}
simplifies to:
\begin{multline}
    \cos(\theta(i\omega^{-1}))=
    \big(x_1(i\omega^{-1})-
    x_1((i-1)\omega^{-1})\big)\\
    \cdot \big[(
        x_1(i\omega^{-1})-
        x_1((i-1)\omega^{-1})
    )^2\\
    +
    (
        x_1(i\omega^{-1})
        \sin(\phi_i)-
        x_1((i-1)\omega^{-1})
        \sin(\phi_{i-1})
    )^2
    \big]^{-\tfrac{1}{2}}\\
    = \frac{1}{\sqrt{1+A^2}}
    \label{eq:orient-bound}
\end{multline}
with
\begin{equation}
    A = \frac{x_1(i\omega^{-1})
    \sin(\phi_i)-
    x_1((i-1)\omega^{-1})
    \sin(\phi_{i-1})}{%
    x_1(i\omega^{-1})
    -x_1((i-1)\omega^{-1})}
\end{equation}
and $\phi_i=
x_2(i\omega^{-1})/
x_1(i\omega^{-1})$.

For we want to bound the
orientation change
-- i.e. we want
$\theta(i\omega^{-1})
<\epsilon$ for $\epsilon$
small enough --,
we want to find a lower
bound for its cosine; i.e.
we want to lower bound
$(1+A^2)^{-1/2}$
in~\eqref{eq:orient-bound}.
However, $A$ depends on what
is the angle of the
$i\omega^{-1}$ position
w.r.t the horizontal, which
relates to the angle
shift $\theta(i\omega^{-1})$
we are looking for.
Such angle depends on
the initial setting of the
ODE system, namely, the
pedestrian density and their
starting locations. Hence,
we bound~\eqref{eq:orient-bound}
for every setup of the ODE.

To overcome such limitation,
we approximate
$\theta(i\omega^{-1})$
considering that pedestrians
are distributed as a PPP,
and checking their average
influence on the pedestrian
of interest
--
see \S\ref{subsec:ppp}.

\section{Radius for
Influence Isolines}
\label{app:radius}

As illustrated in
Fig.~\ref{fig:isolines},
the radius for the influence
isoline $h(r)s(\theta)=H$
changes with $\theta$.
Namely, the isoline
$H$ is fully determined by
$r,\theta$ with the following
expression~\cite{pedestrians}
\begin{equation}
    H= \begin{cases}
        \frac{p\cdot e^{\frac{1}{(r/R_h)^{2}-1}}}{     1+e^{-(\cos(\kappa\theta)-x_0)/R_s}   }, & \left(\frac{r}{R_h}\right)^2<1\\
        0, & \left(\frac{r}{R_h}\right)^2\geq1
    \end{cases}
\end{equation}
with $p=3.59,R_h=0.7,\kappa=0.6,R_s=0.03,
x_0=0.3$~\cite{pedestrians}.

Hence, it is possible to
obtain to reverse the
above equation to know
the isoline radius $r$
w.r.t the angle $\theta$
and isoline value $H$:
\begin{equation}
    r(\theta,H)=
        R_h
        \sqrt{%
        1+\frac{1}{%
        \log\left(
            \frac{H}{p}
        \left(1+e^{-(\cos(\kappa\theta)-x_0)/R_s}
        \right)\right)}
        }
    \label{eq:radius}
\end{equation}
if 
$1/\log(\cdot) \in (-1,0)$,
and zero otherwise.

\section{Correction
in 802.11p Node Model}
\label{app:correction}

According to~\cite{BaiocchiAoI},
the PDR depends on the transmission
probability $\tau$ of an 802.11p node
-- see~\cite[(15)]{BaiocchiAoI}.
$\tau$ depends on
the probability of having zero
packets in the queue of the 802.11p device
$\pi_0$. Such probability $\pi_0$
depends~\cite[(12)]{BaiocchiAoI}
on the Laplace transform of the
time it takes to serve/send a packet,
a random variable denoted as $C$.
The service time $C$ is defined
in~\cite[(3)]{BaiocchiAoI} as
\begin{equation}
    C=\sum_{j=1}^K X^{(j)} + \delta + T_0
\end{equation}
with $K$ \emph{a discrete random variable
uniformly distributed in the set
$\{i,\ldots,W_0\}$}, and $W_0$ the maximum
size of the 802.11p contention window.

Consequently the Laplace transform of
$C$ is obtained as
\begin{multline}
    \mathcal{L}_C(s)=\mathbb{E}_C[e^{-s c}]
    =\sum_{j=1}^{W_0}\mathbb{E}_C[e^{-s c}|\
    K=k]\frac{1}{W_0}\\
    =\frac{e^{-(\delta+T_0)s}}{W_0}\sum_{k=1}^{W_0}
    \int_0^\infty e^{-s \sum_{j=1}^k x^{(j)}}
    f_{C-\delta-T_o}\left(\sum_j^k x^{(j)}
    \right)\ dc'\\
    =\frac{e^{-(\delta+T_0)s}}{W_0}\sum_{k=1}^{W_0}
    \prod_{j=1}^k\int_0^\infty 
    e^{-s x^{(j)}}
    f_{X}(x^{(j)})\ dc\\
    =\frac{e^{-(\delta+T_0)s}}{W_0}\sum_{k=1}^{W_0}\mathcal{L}_{X}^k(s)
    =\frac{e^{-(\delta+T_0)s}\left[1-\mathcal{L}_X^{W_0+1}(s)\right]}{W_0 \left[1-\mathcal{L}_X(s)\right]}
    \label{eq:laplace}
\end{multline}
with $\mathcal{L}_X(s)$ the Laplace transform
of a virtual slot in
802.11p~\cite[(6)]{BaiocchiAoI}.

Note that~\eqref{eq:laplace} differs
from~\cite[(7)]{BaiocchiAoI} because the
numerator has the Laplace transform of
$X$ to the power of $W_0$ rather than
$W_0+1$. Such sublte typo in the manuscript
leads to smaller PDR. That is, with our
correction the model slightly increases the
802.11p PDR.

\begin{IEEEbiography}[{\includegraphics[width=1in,
height=1.25in,clip,
keepaspectratio]{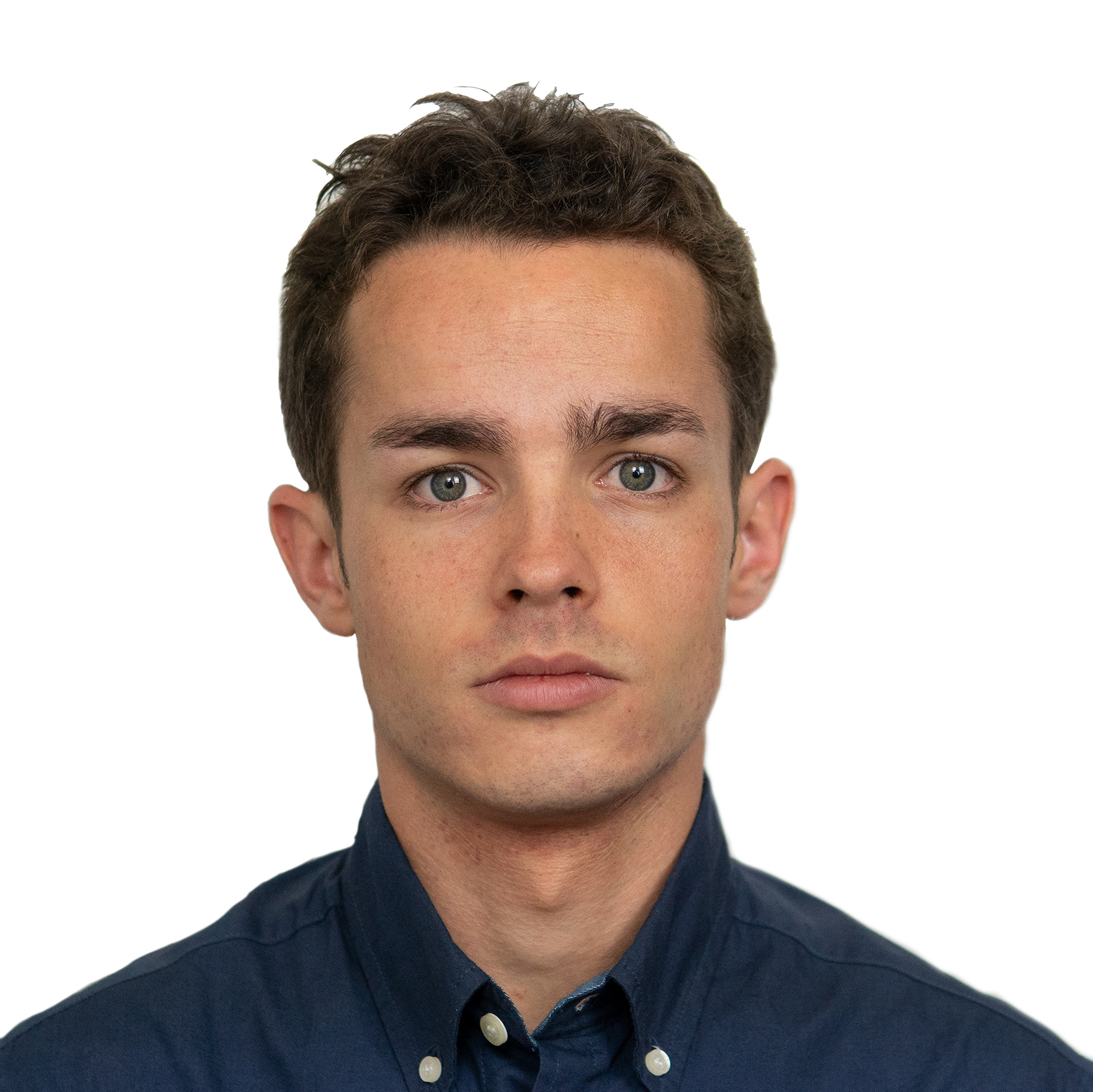}}]{Jorge Martín Pérez}
is an assistant professor at the Universidad
Politécnica de Madrid (UPM), Spain.
He obtained a B.Sc in mathematics,
and a B.Sc in computer science, both
at Universidad Autónoma de Madrid (UAM) in
2016. He obtained his M.Sc. and Ph.D in
Telematics from Universidad Carlos III de
Madrid (UC3M) in 2017 and 2021, respectively.
Jorge worked as postdoc at UC3M (until 2023)
in national and EU funded projects.
His research focuses in optimal resource
allocation in networks.
\end{IEEEbiography}

\begin{IEEEbiography}[{\includegraphics[width=1in,
height=1.25in,clip,
keepaspectratio]{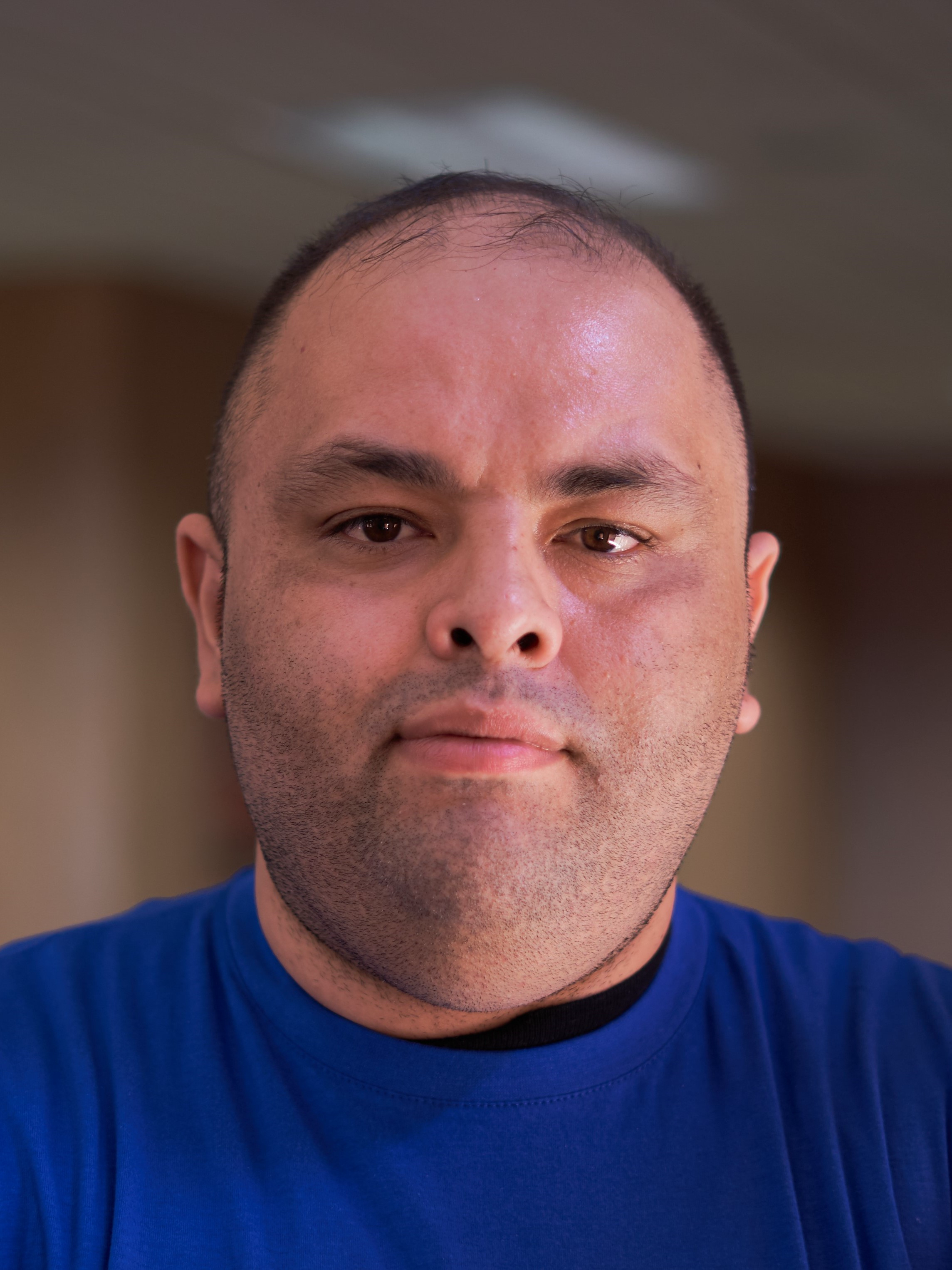}}]{Oscar Amador Molina}
is an assistant professor at Halmstad University in Sweden. He obtained a B.Sc. in Telematics Engineering at Universidad Politécnica de Durango, Mexico, in 2012, and the M.Sc. and Ph.D. degrees in Telematics Engineering at Universidad Carlos III de Madrid (UC3M), Spain, in 2016 and 2020, respectively. His research interests include vehicular networking, protection of vulnerable road users, and intelligent transport systems.
\end{IEEEbiography}

\begin{IEEEbiography}[{\includegraphics[width=1in,
height=1.25in,clip,
keepaspectratio]{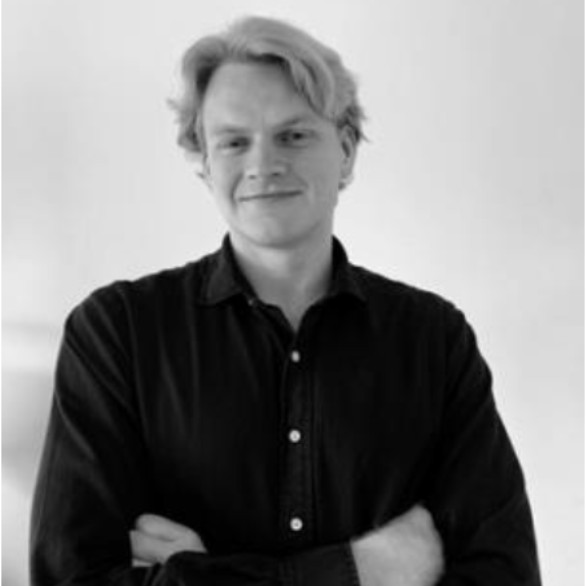}}]{Markus Rydeberg}
obtained an M.Sc. in Data Science and Engineering from the School of Information Technology at Halmstad University in 2023. His master's thesis was on protection of Vulnerable Road Users.
\end{IEEEbiography}

\begin{IEEEbiography}[{\includegraphics[width=1in,
height=1.25in,clip,
keepaspectratio]{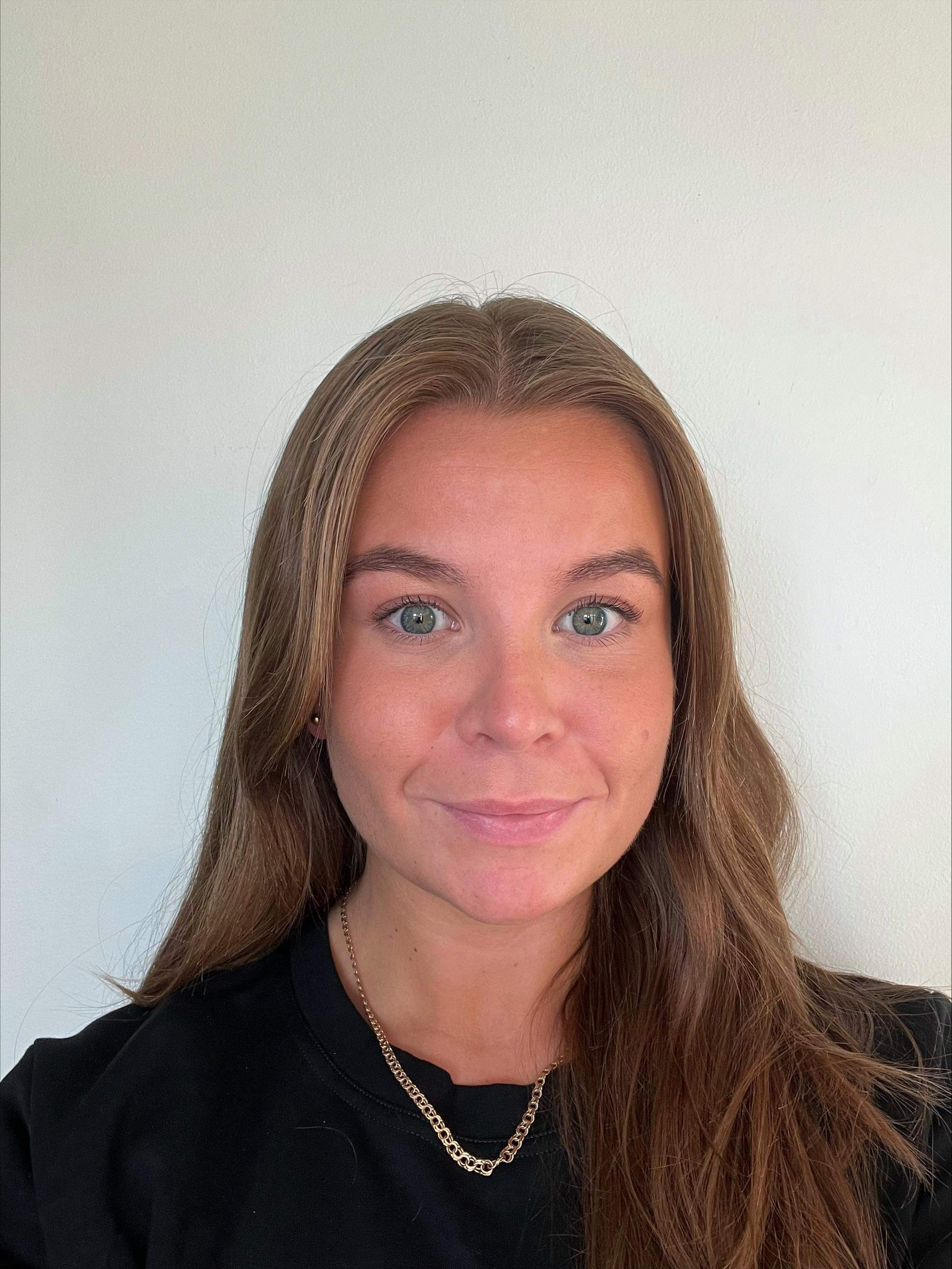}}]{Linnéa Olsson}
obtained an M.Sc. in Data Science and Engineering from the School of Information Technology at Halmstad University in 2023. Her master's thesis was on protection of Vulnerable Road Users.
\end{IEEEbiography}

\begin{IEEEbiography}[{\includegraphics[width=1in,
height=1.25in,clip,
keepaspectratio]{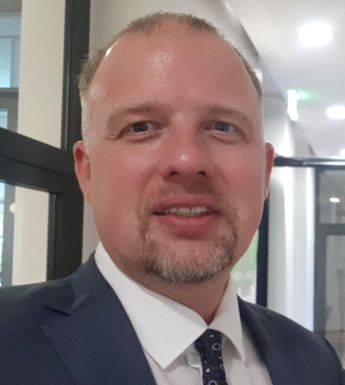}}]{Alexey Vinel}
[SM´12] is a professor at the Karlsruhe Institute of Technology (KIT), Germany. Before he joined KIT in October 2022, he was a professor at the University of Passau, Germany. Since 2015, he has been a professor at Halmstad University, Sweden (now part-time). His areas of interests include vehicular communications and cooperative autonomous driving. He has led several research projects including the Knowledge Foundation synergy project SafeSmart 2019-2024. He received his Ph.D. degree from the Tampere University of Technology, Finland in 2013. He has been a recipient of Alexander von Humboldt Foundation fellowship in 2008.
\end{IEEEbiography}


\end{document}